\documentclass[aps,prx,twocolumn,superscriptaddress]{revtex4-2}
\usepackage{amsmath}
\usepackage{amssymb} 
\usepackage{amsthm}
\usepackage{graphicx}
\usepackage{dcolumn}
\usepackage{bm}
\usepackage{braket}
\usepackage{xcolor}
\usepackage[normalem]{ulem}
\usepackage[hidelinks,colorlinks=false]{hyperref}
\usepackage[capitalise]{cleveref}
\usepackage{algorithm}
\usepackage{algpseudocode}
\newcommand{\abs}[1]{\left\vert#1\right\vert}

\newcommand{\Tr}{\operatorname{Tr}}
\newcommand{\dl}[2]{\ensuremath{\delta_{\{{#1},{#2}\}}}}

\definecolor{blue}{rgb}{0,0,1}

\begin{document}

\title{Computing noise-canceling observables via Pauli propagation}

\author{Andrew Eddins}
\email{aeddins@ibm.com}
\affiliation{IBM Quantum, IBM Research -- Silicon Valley, 555 Bailey Ave, San Jose, CA 95141, USA}
\author{Caleb Johnson}
\affiliation{IBM Quantum, IBM T.J. Watson Research Center, Yorktown Heights, NY 10598, USA}
\author{Alberto Baiardi}
\author{Francesco Tacchino}
\affiliation{IBM Quantum, IBM Research Europe -- Zurich, 8803 Rüschlikon, Switzerland}
\author{Ewout van den Berg}
\author{Roy Elkabetz}
\affiliation{IBM Quantum, IBM T.J. Watson Research Center, Yorktown Heights, NY 10598, USA}
\author{Vinay Tripathi}
\affiliation{IBM Quantum, Research Triangle Park, NC 27709, USA}
\author{Swarnadeep Majumder}
\affiliation{IBM Quantum, IBM Research Cambridge, Cambridge, MA 02139, USA}
\author{Max Rossmannek}
\affiliation{IBM Quantum, IBM Research Europe -- Zurich, 8803 Rüschlikon, Switzerland}
\author{Liran Shirizly}
\author{Abhinav Kandala}
\affiliation{IBM Quantum, IBM T.J. Watson Research Center, Yorktown Heights, NY 10598, USA}

\date{\today}

\begin{abstract}
The pursuit of quantum advantage is driving the co-evolution of quantum processors and classical simulation methods. Despite advances in scale and quality, the accuracy of quantum simulation is ultimately limited by error rates and sampling overheads. Similarly, while classical simulation methods such as Pauli propagation have made remarkable progress, their accuracy is ultimately limited by the exponential growth of operator paths and the truncations needed to control memory and runtime. Here we show that these complementary limitations can be mitigated by embedding Pauli propagation within a hybrid error-mitigation framework that reduces quantum sampling overhead while achieving lower truncation errors with fewer classical resources than traditional Pauli propagation alone. In this framework, a target observable is classically propagated through noise-canceling inverse channels, producing a modified observable that is measured directly on a quantum processor. We prototype two implementations and benchmark their performance numerically on canonical models that challenge traditional Pauli propagation. We also perform experiments on a quantum processor using 56 superconducting qubits, revealing the tradeoffs of their respective truncation strategies. These results illustrate how classical and quantum resources can be orchestrated to extend observable estimation beyond the limits of either approach alone, providing a foundation for quantum-centric supercomputing and future demonstrations of quantum advantage.

\end{abstract}

\maketitle

Accurately estimating observable properties of shallow quantum circuits is a key building block of several near-term quantum algorithms~\cite{Preskill2018nisq,bharti2022nisq}, with many scientific applications such as chemistry, high-energy physics, and parameter optimization, and itself remains a leading candidate for demonstrating useful quantum advantage. The drive towards advantage has spurred a race between development of quantum and classical simulation capabilities. On the classical side, Pauli propagation~\cite{rall2019pauliprop, rudolph2023,begusic2025cpt, rudolph2025pauliprop} has recently emerged as a powerful method for simulating quantum expectation values, demonstrating competitive performance against both leading classical simulation techniques and quantum experiments. Pauli propagation evolves an observable through the circuit by tracking its decomposition into Pauli operators. However, under sufficiently complex dynamics, the exponential proliferation of Pauli paths necessitates truncation, degrading the accuracy of the final expectation value.

On the quantum side, advances in hardware and error mitigation have enabled accurate observable estimation at scales beyond brute-force classical simulation~\cite{Kim2023utility}. 
Quantum error mitigation can reduce or remove noise-induced bias in measured expectation values, allowing noisy processors to produce reliable results even before fault tolerance~\cite{temme2017mitigation,benjamin2018efficient}. However, unbiased mitigation methods such as probabilistic error cancellation (PEC) incur sampling costs that grow exponentially with circuit noise, limiting tractable circuit volumes~\cite{temme2017mitigation,berg2023pec}. 

These limitations suggest a natural hybrid strategy that leverages ever-developing classical simulation techniques to supplement quantum computation and extend to circuit volumes that are beyond the individual limitations of either approach. Here, we realize this strategy by using Pauli propagation to construct noise-canceling observables that are subsequently measured on quantum hardware. We introduce two implementations, propagated noise absorption and Clifford-Dyson error mitigation, respectively truncating greedily or perturbatively, and test the methods on two-dimensional transverse-field Ising Trotter circuits; similar circuits were recently found to be challenging for direct classical simulation via Pauli propagation \cite{haghshenas2025digital}. Numerical simulations provide cases where these methods outperform the use of Pauli propagation to simulate the noiseless expectation value directly, while also reducing the sampling overhead of standard probabilistic error cancellation. Finally, a 56-qubit experiment on quantum hardware demonstrates the practical viability of the methods at non-trivial scales. These results add to the growing body of work on the use of such hybrid, quantum-centric supercomputing workflows~\cite{seelam2026} to extend the reach of error mitigation~\cite{eddins2024lightcone,filippov2023tem,fischer2026,majumder2026} and near-term quantum computing~\cite{fuller2025obp,robledo2025chemistry,Robertson2025,Robertson2025b}.

\section{Computing a noise-canceling observable}

\begin{figure*}[t!]
    \includegraphics[width=0.75\linewidth]{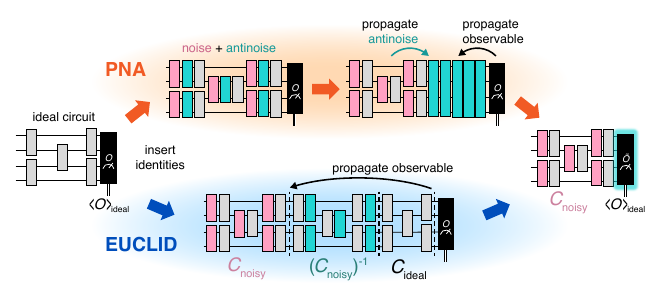}
    \caption{\textbf{Classically propagated noise cancellation schemes.} a) \textit{Propagated noise absorption} (PNA) transforms the ideal circuit into an equivalent pairing of the noisy hardware gate layers combined with a modified, noise-canceling observable. The noisy gate layers are produced by classically propagating equal and opposite antinoise to the end of the circuit. Propagating the observable backwards through this evolved antinoise embeds noise cancellation into the resulting observable $\tilde{O}$, such that estimating $\langle\tilde{O}\rangle$ on the noisy hardware produces the ideal expectation value. b) \textit{Euclid} constructs $\tilde{O}$ by backpropagating $O$ through a noise deconvolution map consisting of the inverse noisy circuit followed by the ideal circuit. The near-mirror structure of the map facilitates propagation via Clifford perturbation theory (CPT), a systematic expansion into Pauli paths ordered by the number of gates acting nontrivially on each path.}
    \label{fig_overview}
\end{figure*}

We model a noisy circuit $\mathcal{C}_\text{noisy}$ as a sequence of operations time-ordered by the index $t$, where each operation may be either a unitary gate $U_t$ or a noise channel $\Lambda_t$. We assume all $\Lambda_t$ are Pauli channels, as in circuits compiled with Pauli twirling \cite{PhysRevLett.76.722,knill_twirling, wallman2016randcomp}, generated by composition of local, independent Pauli Lindblad channels $\Lambda_{t,i}$ \cite{berg2023pec}:
\[
\Lambda_t(\cdot) = \bigcirc_{i}\Lambda_{t,i}(\cdot)=\bigcirc_{i}\big( (1-p_{t,i})(\cdot) + p_{t,i} E_{t,i} (\cdot) E_{t,i}\big).
\]
Here $(\cdot)$ is the channel input, which is conjugated by each local Pauli error $E_{t,i}$ with probability $p_{t,i} = (1-e^{-2\lambda_{t,i}})/2$ for noise strength $\lambda_{t,i}$. The inverse noise map $\Lambda_t^{-1}$ has the same form as $\Lambda_t$, but with negated $\lambda_{t,i}$ parameters; instead of $p_{t,i}$ we write $q_{t,i} \leq 0$ to emphasize it is a quasiprobability. Although such ``antinoise'' is nonphysical -- a map but not a channel -- it can be realized in expectation via quasi-probabilistic sampling~\cite{berg2023pec}. In PEC, one thereby inserts antinoise maps $\Lambda_t^{-1}$ by each noise channel, leaving behind noiseless operations, e.g. $\Lambda_{t+1}^{-1}\Lambda_{t+1}\mathcal{U}_t = \mathcal{U}_t$, where $\mathcal{U}_t$ is the channel form of $U_t$.

Rather than including the inverse maps in the quantum circuits, we insert them only conceptually before embedding them in the observable via Pauli propagation. Supposing a basis gate set of Pauli rotations, forward evolution of a Pauli $P$ past a single gate $U_t = \exp(-iR_t (\theta_t/2))$ is
\begin{equation}
    U_t P U_t^\dagger = \left(\cos(\theta_t) -i\sin(\theta_t)R_t\right)^{\langle P, R_t \rangle} P,
    \label{eq:evolve_gate}
\end{equation}
where 
$\langle P, R_t \rangle$ is 0 (1) if $P$, $R_t$ (anti-)commute. Similarly, evolution of a Pauli through 
an 
inverse map $\Lambda^{-1}_{t,i}$ (forward or backward) is
\begin{equation}
  \Lambda^{-1}_{t,i}(P)
    = (1 -2 q_{t,i})^{\langle P, E_{t,i} \rangle} P.
    \label{eq:evolve_noise}
\end{equation}
We apply these propagation rules classically to transform the ideal circuit $\mathcal{C}_\text{ideal} = \bigcirc_t\mathcal{U}_t$ and observable $O$ into a noisy circuit $\mathcal{C}_\text{noisy}$ and modified observable $\tilde{O}$, which when run on the noisy quantum hardware yield the desired ideal expectation value $\langle O \rangle$.
In practice, we perform a classical computation to remove antinoise from the bulk of the circuit by absorbing it into $\tilde{O}$. 
Below we consider two propagation methods, ``Propagated Noise Absorption'' (PNA) and ``Error mitigation Using CLIfford-Dyson'' (Euclid), which compute $\tilde{O}$ using different approximations and orders of operations (Fig.~\ref{fig_overview}). 
PNA starts with all terms in the error-propagation problem and greedily truncates small terms until the problem fits onto available classical memory. In contrast, Euclid builds up terms systematically in their Pauli-path expansions in the style of Clifford Perturbation Theory (CPT)~\cite{begusic2025cpt}. The expectation value of $\tilde{O}$ obtained via PNA or Euclid is then estimated on the noisy QPU directly. A similar approach to absorb noise in a modified observable using tensor networks was reported in \cite{filippov2023tem,fischer2026}.

\subsection{Propagated Noise Absorption}
PNA evolves antinoise forwards to the end of the circuit, and evolves the observable backwards through this antinoise, as shown in Figure~\ref{fig_overview} . These two computations may proceed concurrently. Both steps will in general have exponential complexity if performed exactly, so are approximated by truncating small terms. Our PNA implementation is currently available as an open-source Qiskit Add-On \cite{pna_addon}.

Starting at low $t$, we insert the noise-antinoise pair $\Lambda_t^{-1} \Lambda_t$ in the ideal circuit, and propagate each $\Lambda_{t,i}^{-1}$ (but not $\Lambda_{t,i}$) forward to the end of the circuit one by one before repeating for the next noise-antinoise pair. This ordering ensures propagation is through only unitary gates, avoiding problematic correlations introduced when evolving one map through another (see App. \ref{app:lightcones}). The map $\Lambda^{-1}_{t,i}$ applies Pauli error $E_{t,i}$ with quasiprobability $q_{t,i}\leq0$. It may be propagated to the end of the circuit by noting that the time-evolved map simply applies the time-evolved error,
\begin{equation}
\label{channel_prop}
    U\Lambda^{-1}_{t,i}U^\dagger(\cdot) = (1-q)(\cdot) + q(UEU^\dagger)(\cdot)(U^\dagger EU),
\end{equation}
suppressing the subscripts on $E$ and $q$ for brevity. Combining this with Eq.~\eqref{eq:evolve_gate} and repeatedly applying for all gates following the antinoise at time $t$, we obtain the propagated error $E'$ as a weighted sum of Paulis. We use each $E_{t,i}'$ to update the observable in the Heisenberg picture,
\begin{equation}\label{eq:applyantinoise}
    \Lambda_{t,i}'^{-1}O = (1-q)O + qE'OE',
\end{equation}
cumulatively producing the final modified observable $\tilde{O}$. Note that Eq.~\eqref{eq:evolve_noise} does not apply as the evolved error is no longer a Pauli.

Propagation of $E$ forward and of $O$ backward both
involve branching Pauli terms growing exponentially in memory with the number of non-Clifford gates. To regulate this complexity, each computation is approximated by truncating intermediate results, keeping at most the $K_E$ or $K_O$ largest Pauli terms, respectively, making memory and runtime requirements predictable and avoiding the allocation overhead of freely growing arrays. Lower values of $K_E$ and $K_O$ may suffice in circuits where the Pauli decompositions are dominated by relatively few terms, such as circuits comprised of near-Clifford gates. We also truncate terms with coefficient magnitudes below a threshold $\varepsilon$. Working with scaled operators $\sqrt{|q_{t,i}|}E_{t,i}$ accelerates truncation since $|q_{t,i}|$ is typically small. To avoid the temporary creation and truncation of very large numbers of terms, backpropagation is performed by directly computing only the largest $K_O$ terms (App.~\ref{app:zigzag}). The result of backpropagating $O$ through all $\Lambda_{t,i}'^{-1}$ is $\tilde{O}$.

\subsection{Euclid}
In Euclid, $\tilde{O}$ is computed by approximately applying the deconvolution map $\mathcal{D} =(\mathcal{C}_\text{ideal}\ \mathcal{C}_\text{noisy}^{-1})^\dagger$ on the target observable, as originally proposed in ~\cite{filippov2023tem,mangini2022qubit} (Fig.~\ref{fig_overview}), recovering the ideal expectation value,
\begin{equation}
  \langle O \rangle 
    = \Tr \Big[ O\ \mathcal{C}_\text{ideal}\ \mathcal{C}^{-1}_\text{noisy}\ \mathcal{C}_\text{noisy} |0 \rangle\langle 0| \Big]
    = \Tr[\mathcal{D} O \rho]
\end{equation}
where $\rho = \mathcal{C}_\text{noisy}|0 \rangle\langle 0|$ is prepared by the noisy QPU. Although backpropagation of $O$ through the original circuit is generally prohibitively difficult, propagation through $(\mathcal{C}_\text{ideal} \ \mathcal{C}_\text{noisy}^{-1})^\dagger$ can be easier due to the approximate cancellation of gates with their inverses. Euclid leverages this structure to provide a perturbative expansion of $\mathcal{D}$ for arbitrary quantum circuits.

\begin{figure}[!t]
  \centering
  \includegraphics[width=.95\linewidth]{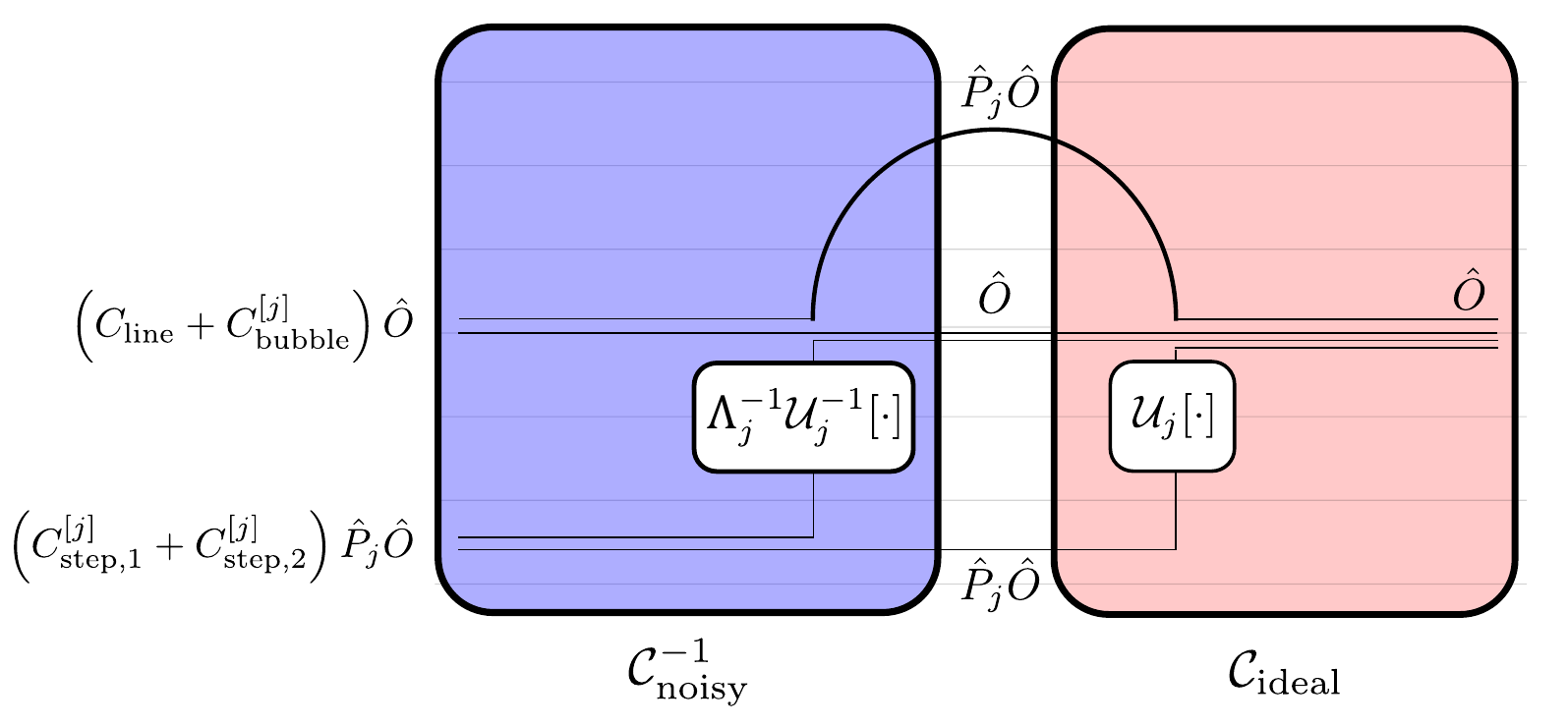}
  \caption{
  \textbf{Euclid diagrams.}
  The inverse noisy circuit $\mathcal{C}_\text{noisy}^{-1}$ is represented as a blue box, while the ideal circuit $\mathcal{C}_\text{ideal}$ is represented as a red box. The line diagram 
  indicates
  the cosine branch 
  is taken for all gates,
  yielding a backpropagated Pauli equal to the original Pauli observable, $O$.
  The bubble diagram indicates the sine branch is taken for the $j$th gate in both $\mathcal{C}_\text{noisy}^{-1}$ and $\mathcal{C}_\text{ideal}$, again yielding Pauli $P_j P_j O = O$, but with a different coefficient.
  In the step diagram, the sine branch is taken either in $\mathcal{C}_\text{noisy}^{-1}$ or in $\mathcal{C}_\text{ideal}$. The final operator is $P_j O$, and the operator after the application of $\mathcal{C}_\text{ideal}$ is $O$ ($P_j O$) if the sine branch is taken in $\mathcal{C}_\text{noisy}^{-1}$ ($\mathcal{C}_\text{ideal}$).}
  \label{fig:Euclid-Diagrams}
\end{figure}

In the spirit of CPT~\cite{begusic2025cpt, begusic2024fast}, we expand $\tilde{O} = \mathcal{D}O$ as a sum of single-Pauli paths, or diagrams, in the backpropagation of $O$ through $\mathcal{D}$.
We classify diagrams in terms of lines, steps, and bubbles (Fig.~\ref{fig:Euclid-Diagrams}).
A step indicates transformation from one Pauli to another, associated with a factor of $\sin\theta_t$. A bubble, or pair of steps centered horizontally, adds a correction to an existing Pauli term. $\tilde{O}$ can be iteratively grown in numbers of steps, introducing more terms, and bubbles, improving coefficient accuracy. These diagrams help to collect all terms with up to $K$ sine factors, which can be made small by expressing the circuit in the Clifford interaction picture, such that all $\left| \theta_t \right| \leq \pi/4$ and, hence, $\abs{\sin(\theta_t)} \leq \frac{1}{\sqrt{2}}$. Further details and examples appear in {Appendix~\ref{sec:euclid_theory_app}}.

\subsection{Estimation of $\tilde{O}$ on the QPU}\label{sec:estimation}
Estimating the $\tilde{O}$ obtained via PNA or Euclid on the noisy QPU provides the mitigated expectation value $\langle O\rangle$. The structure of $\tilde{O}$ determines the sampling cost, which multiplies the total QPU time needed. As the sampling considerations are largely independent of the method used to compute $\tilde{O}$, analyses done in the context of tensor-error mitigation broadly apply \cite{filippov2023tem, filippov2024scalability}. Sampling cost reductions derive primarily from two sources: individual errors may partially commute with the observable, and, at a risk of introducing bias, some noise terms may be truncated and thus not mitigated. At the same time, the possibility of exponentially many Pauli terms in $\tilde{O}$ can make measurement more challenging. When relatively few terms dominate $\langle\tilde{O}\rangle$, as expected for circuits with sufficiently low noise (small $\lambda_t$) or low magic (small $\sin\theta_t$ or $\cos\theta_t$), measuring only terms with the largest coefficient magnitudes can provide acceptable accuracy, effectively truncating all but $K_M \geq 1$ terms. Provided $O$ consists of Pauli terms of commensurate coefficients, these original Paulis likely comprise the largest terms in $\tilde{O}$ as well, but with modified coefficients, and measuring only these original Paulis can be sufficient \cite{filippov2024scalability}. 
In our experimental demonstration, we measured only the original Paulis in $O$, though our numerics show this approach is not well suited for all problems. Many grouping and weighting strategies have been developed to facilitate estimation of many-term observables \cite{huang2020shadows, hadfield2020lbcs, garciaperez2021adaptiveicm, fischer2024icm, huang2021derandom, gresch2025shadowgroup}, which may be utilized to increase $K_M$ or otherwise improve accuracy. Additional sampling-cost details appear in App.~\ref{app:samplingcosts}.

\section{Numerical examples}\label{sec:numerics}

\begin{figure}[!t]
    \centering
    \includegraphics[width=0.99\linewidth]{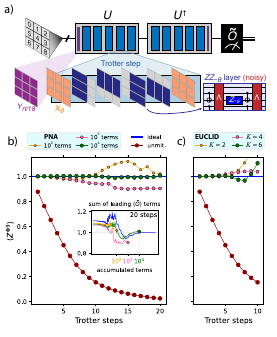}
    \caption{\textbf{9-qubit numerics.} a) A 3-by-3 square lattice of qubits is evolved first forwards, then backwards in time under the second-order Trotterization of the transverse-field Ising Hamiltonian. A parameter $\theta$ sets the $R_X$ (orange) and $R_{ZZ}$ (blue) gate angles. CNOT gates are transpiled to $H$-conjugated CZ primitive gates (not shown). Each two-qubit gate layer is followed by the action of one of four unique Pauli-Lindblad noise models (red). b,c) Expectation values $\langle Z^{\otimes9}\rangle$ versus number of Trotter steps in $U$ (first half of circuit) with $\theta = 0.05$, for (b) PNA and (c) Euclid. All truncation settings greatly reduce bias compared to unmitigated results, with the highest settings retaining accuracy to larger depths. Insets discussed in text.}
    \label{fig:easy_numerics}
\end{figure}

To study the accuracy of these approximate methods, we numerically simulate mitigation by PNA and Euclid of 
noisy Trotter evolution under the transverse-field Ising model Hamiltonian. Each Trotter step consists of interleaved $X$ and $ZZ$ rotations. 
We consider the mirror circuits in Fig.~\ref{fig:easy_numerics}(a) for a 3-by-3 square lattice of qubits, which evolve several steps forwards then backwards, such that the noiseless circuit measures the original state $\ket{0}$, with $\langle O \rangle = 1$ for any diagonal $O$. Each entangling layer is followed by a sparse Pauli-Lindblad noise model with generators $E_{t,i}\in\{X,ZZ\}$ and rates randomly chosen with a target mean error-per-layered-gate (EPLG) \cite{mckay2023eplg} of 0.0025, comparable to EPLG reported on recent IBM Heron processors.

Figures~\ref{fig:easy_numerics}(b,c) show numerically-simulated performances of PNA and Euclid on the noisy circuit with target observable $O = Z^{\otimes9}$. Small gate angles 
with $\theta = 0.05$, corresponding to a small Trotterization timestep, make this pedagogical example easy for Pauli propagation methods. PNA was run for three truncation settings, $K_E = K_O \in [100, 1000, 10000]$ terms, and Euclid was run at orders $K\in[2,4,6]$. For each $\tilde{O}$ produced, the expectation value was evaluated with respect to the exact density matrix at the end of the noisy circuit; free of shot noise, any deviations from the ideal value of 1 indicate truncation bias. All the considered settings significantly reduce bias compared to the unmitigated result. However, visible bias accumulates after a small number of Trotter steps for more heavily truncated approximations. Keeping more terms suppresses overall bias and slows fluctuations thereof, with the largest PNA computation providing the greatest accuracy. Focusing on that 20-step circuit, the PNA inset indicates convergence of each mitigation result as more Pauli terms are included in $\langle\tilde{O}\rangle$, simulating the effect of neglecting to measure smaller Pauli terms on the noisy QPU. In this example, measuring only the original term in $\tilde{O}$ would lead to a bias of approximately $+0.1$, regardless of other truncation settings, motivating incorporating more advanced methods to estimate many-term observables in future development.

\begin{figure}[!t]
  \centering
  \includegraphics[width=0.9\linewidth]{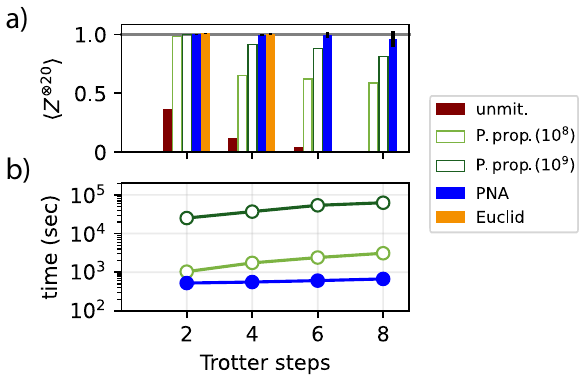}
  \caption{\textbf{20-qubit numerics.}  Numerically-simulated results for a circuit similar to that in Fig.~\ref{fig:easy_numerics}a on a 4-by-5 qubit grid. PNA (max terms=$10^4$) or Euclid ($K=6$) largely eliminate the bias present in the unmitigated result. The PNA computation, including $10^6$ simulated QPU shots at $2\,\text{kHz}$ sampling frequency, was both faster and more accurate than a direct computation of the noiseless circuit via Pauli propagation.}
  \label{fig:4x5numerics}
\end{figure}

We now consider another numerical example of a 4-by-5 qubit lattice, but with a larger gate-angle parameter $\theta=0.25$, that is more challenging for traditional Pauli propagation methods, and compare it to the performance of PNA and Euclid for the estimation of the full weight-20 $Z$ observable (see in Figure~\ref{fig:4x5numerics}). 
The chosen noise models include all two-local Pauli generators with randomized rates targeting an EPLG of approximately 0.001, and a maximum circuit depth of 8 trotter steps. At this scale, the exact density matrix is not readily available, so shot noise limits precision. $10^6$ noisy shots were simulated to calculate unmitigated expectation values. 

PNA was run with $K_E = K_O = 10^4$ for each setting. At all circuit depths, PNA mitigation removes the bias up to shot-noise uncertainty. In contrast, we see that observable back-propagation of the noiseless circuit produces larger bias than the PNA results, at all the considered depths, even while propagating up to $10^9$ terms. We estimate the time to solution for the PNA experiments, including both the classical runtime and the QPU time assuming a realistic sampling rate of 2 kHz. Even with much more time and memory (Fig.~\ref{fig:4x5numerics}b), the purely-classical observable backpropagation routine produced results with much larger bias than the simulated PNA results, for all four circuit depths tested. Despite the ease of exact classical simulation at this scale, these numerics present an optimistic view on the use of such hybrid workflows, and motivate experimental tests at scale.

\section{Experimental results} 
\label{sec_experiments}
We finally tested the methods in practice on a 2D grid of 56-qubits of \texttt{ibm\_boston}, an IBM Heron processor with a reported EPLG of approximately $2\times10^{-3}$ and employing a novel frame randomization to suppress crosstalk in calibrations (see Appendix~\ref{app:random_frames}). 
These tests also used mirror circuits with Trotterized transverse field Ising dynamics, now on a heavy-hex lattice to fit the device connectivity. At a fixed depth of four Trotter steps for the forward evolution (totaling 48 Pauli-twirled CZ layers in the entire circuit), we swept the angle $\theta$ in the $R_X(\theta)$ and $R_{ZZ}(\theta)$ layers between 0 and $\pi$/2. To more cleanly study mitigation as a function of Clifford-ness, we omit the initial small $R_Y$ rotation, and also the second $R_X$ rotation in each step.

Recent improvements in user-facing \cite{samplomatic} and backend software components facilitated the fast execution of Pauli-twirled circuits. The standard noise-learning workload executed in 12.7 minutes of Qiskit-Runtime usage, providing the noise model used by PNA in postprocessing, while the main experiment workload executed in 27.0 minutes of usage. The latter included all nine $\theta$ values, each comprised of 100,000 circuit executions (2,000 twirl re-randomizations at 50 shots each). Additional measurements followed each circuit execution to postselect against leakage out of the computational subspace \cite{Kim2026postselect}, keeping approximately 70\% of the data.

We targeted two observables: the spatial average of all weight-one $Z$ operators as a measure of ensemble qubit performance, and a non-local weight-two operator $Z_{88}Z_{118}$ as an example of a single-term observable with a light cone encompassing a larger number of qubits, both obtained by measuring all qubits in the computational basis. Plotted values include readout-error mitigation via TREX \cite{trex}, with a qubit-averaged single-site correction factor of $1.011(6)$. The unmitigated results deviate from the ideal value of 1 as $\theta$ increases, with larger gate angles causing faster operator spreading that increases spatial overlaps of errors with the observable. 

Mitigation reduces bias at all points. At Clifford points $\theta = 0, \pi/2$ there is no truncation error, and residual error provides a measure of noise model inaccuracy. In contrast, PNA applied to noisy Pauli-propagation simulations, using the same noise model for simulation and mitigation, exactly recovers the ideal value of $+1$ at the Clifford points. For non-Clifford circuits, including more terms in the PNA computations of $\tilde{O}$ reduces significant truncation bias in the domain $\pi/4<\theta<\pi/2$. Although $\theta=\pi/4$ maximizes the classical difficulty of processing an individual gate, truncation bias peaks at a larger angle because the faster operator spreading increases sensitivity to more gates and errors. Truncation error also includes the effect of measuring only those Paulis originally in $O$, discussed in Sec.~\ref{sec:estimation}.

\begin{figure}
    \centering
    \includegraphics[width=0.99\linewidth]{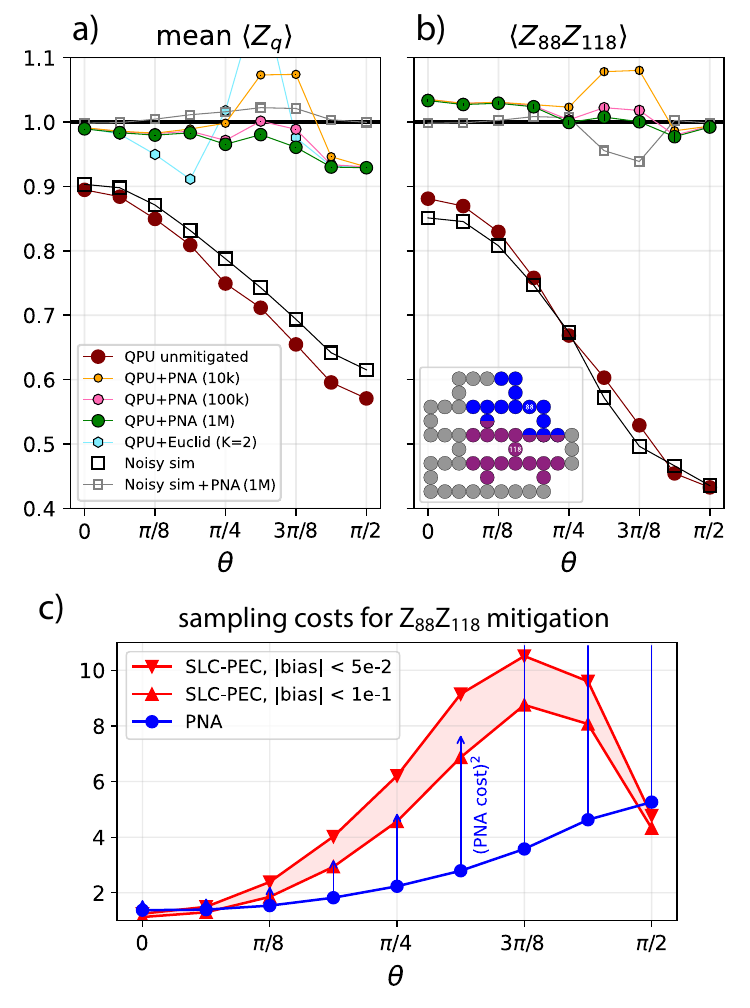}
    \caption{\textbf{Mitigation of 56-qubit experiments.} Raw and mitigated expectation values of a) site-averaged magnetization $\sum_q^N Z_q / N$ and b) $Z_{88}Z_{118}$, measured as a function of gate angle $\theta$ (see Sec.~\ref{sec_experiments} text). Right inset shows qubit connectivity, highlighting the nominal four-step lightcones of the $Z_{88}$ (blue) and $Z_{118}$ (purple) measurements. 
    The ideal expectation values of these mirror circuits are all $+1$ (black horizontal line). 
    Mitigation reduces error at all points. Including more terms when computing $\tilde{O}$ reduces truncation bias.
    Comparison with $10^8$-term Pauli-propagation simulations of the noisy circuit (squares) supports attributing residual error in the PNA-mitigated experiments to noise-model inaccuracy.
     All points approximate $\tilde{O}$ by its projection onto the Paulis in the original observable (Sec.~\ref{sec:estimation}). c) For $Z_{88}Z_{118}$, PNA exhibits a quadratic sampling speedup compared to shaded-lightcone accelerated PEC. The lower PNA variance comes at a cost of unbounded bias. At Clifford points, PNA becomes unbiased and the quadratic separation vanishes, with the SLC-PEC costs slightly reduced by the nonzero bias tolerances chosen.}
    \label{fig:pna_expt}
\end{figure}

Lastly, we compare the sampling cost of PNA mitigation to PEC, for the observable $\langle Z_{88}Z_{118}\rangle$, in Figure~\ref{fig:pna_expt}c.
Since we use the leading-term approximation $K_M=1$ (Sec.~\ref{sec:estimation}), the PNA cost is the squared magnitude of the $Z_{88}Z_{118}$ coefficient in $\tilde{O}$, computed with $K_E=K_O=10^6$. This cost increases with $\theta$ from 1.4 to 5.3, increasing with operator spreading. The cost of naive PEC, which eliminates bias by mitigating all noise in the 56-qubit circuit, is $2.8\times10^5$, an irrelevantly larger value. Since PNA provides a biased estimator, we instead compare to PEC allowed some residual bias. We plot the cost of using a classically shaded lightcone (SLC) \cite{eddins2024lightcone} to parsimoniously mitigate just enough errors with PEC to bound the remaining bias below chosen tolerances of 0.05 and 0.1. Note that a lower tolerance demands a higher sampling cost, and the actual bias remaining may be smaller than the tolerance. For a range of non-Clifford $\theta$ values, the square of the PNA sampling cost roughly tracks the square root of the PEC cost, suggesting the expected quadratic speedup persists even compared to an optimized PEC routine, and when using a real-world Pauli-Lindblad noise model. At Clifford $\theta$ values, approximately half of the Pauli errors in the lightcone of the observable commute with it; skipping these accelerates SLC-PEC quadratically, matching the sample-efficiency of PNA. (The bias tolerance makes SLC-PEC slightly cheaper than the exact PNA mitigation). For near-Clifford $\theta$, the SLC-PEC sampling cost transitions between the PNA cost and its square.

\section{Outlook}\label{sec_conclusion}
The above results recommend classical propagation through an inverted noise model as a practical and flexible means of error mitigation, with Pauli propagation a natural choice given the broad suitability of Pauli noise models. 
Numerical results showed this task can be easier than direct simulation of the noiseless circuit via Pauli propagation, one prerequisite for quantum advantage. Numerical and experimental results demonstrated accuracy at scale, and for nonlinear qubit topologies that can be challenging for alternative tensor network approaches, with expected quadratic sampling cost reductions for non-Clifford experiments. Work remains to improve noise learning accuracy, and to increase $K_M$ by more efficient sampling of many-Pauli observables.

Many potential algorithmic improvements remain to explore. 
Instead of parallel propagation of many antinoise unitaries in PNA, all antinoise could be propagated as one superoperator; this effectively doubles the number of qubits, but reduces backpropagation of $O$ to a single step, which may be easier or allow bounding truncation bias.
Synergistically, the current outer-loop parallelization might be moved inside the propagation steps themselves \cite{broers2025parallelprop}, perhaps leveraging GPU resources, if the increased compute power outweighs communication overhead.
Turning from implementation to application, one may first apply PNA or Euclid to compute $\tilde{O}$ canceling noise in only the later part of a circuit, then apply a time-reversed variant to compute a modified initial state, $\tilde\rho$, canceling noise in the earlier part of the circuit, sampling both operators each circuit execution. Thus halving the Pauli propagation depth would greatly reduce the classical workload; a future study could investigate sampling complexity.
Additionally, in deep circuits with sufficient scrambling properties, errors far from either end might be reasonably approximated as depolarizing noise after propagation (App. \ref{app:samplingcosts}), replacing the most difficult propagations with trivial rescaling.
More broadly, we expect such hybrid mitigation strategies are well poised to benefit from advances on either side in the ongoing race between quantum and classical technologies, and can in principle grow exponentially more powerful as early fault-tolerant systems suppress baseline noise rates and propagation complexity.

\begin{acknowledgments}
    We thank Kristan Temme, Minh Tran, Dekel Meirom, Jennifer Glick, Jessie Yu, Bryce Fuller, Samantha Barron, Brad Mitchell, Brendan Saxberg, Alireza Seif, Isaac Lauer, Ian Hincks, and James Raftery for helpful conversations, technical assistance, and programmatic support. This
research was supported by the U.S. Department of Energy, Office of Science, National Quantum
Information Science Research Centers, Quantum Science Center.
\textbf{Competing interests:} Elements of this work are included in patent applications filed by the International Business Machines Corporation with the US Patent and Trademark Office.
\end{acknowledgments}

\appendix

\section{Sampling costs of estimating $\tilde{O}$}\label{app:samplingcosts}
Whereas PEC attempts to cancel the noise itself, $\tilde{O}$-based methods cancel only the perturbation of the observable by the noise; this approach of treating the problematic symptom rather than curing the underlying disease tends to be easier, manifesting in lower sampling costs. Per Eq. \eqref{eq:applyantinoise}, an antinoise error perturbs $O$ by an amount proportional to their commutator,
\begin{equation}\label{eq:applyantinoise_commutator}
    \Lambda'^{-1}(O)=(1-q)O+qE'OE'=O+q[E',O]E'.
\end{equation}
If this commutator is small, the $\tilde{O}$ sampling cost to mitigate $\Lambda$ decreases, but the standard PEC cost, $\gamma^2_{PEC}=(1-2q)^2$, does not. This benefit compounds exponentially with the number of errors in the circuit, enabling much faster mitigation compared to PEC \cite{filippov2023tem, filippov2024scalability}.

A notable exception is circuits with only Clifford gates, where all relevant errors have maximal commutators ($\{E',O\}=0$), such that both PEC and $\tilde{O}$-methods can obtain the same optimal sampling cost. In the notation of PNA, each forward-evolved $E'_{t,i}$ is a single Pauli. One can always efficiently identify the Paulis $E'_{t,i}$ that commute with $O$, and remove these inconsequential errors from the noise model \cite{eddins2024lightcone}, such that all remaining errors anticommute with $O$. This will eliminate the benefit in Eq. \eqref{eq:applyantinoise_commutator} of the sampling cost depending on the commutator. All $qE_{t,i}'OE_{t,i}'=-qO$ have the same sign so sum constructively to $\tilde{O} = O\prod_{t,i}(1-2q_{t,i})$. The $\tilde{O}$ sampling cost is the square of this rescaling factor, which exactly equals the PEC sampling cost. Thus, when mitigating Clifford circuits, we do not expect $\tilde{O}$-based methods to provide a sampling cost advantage compared to PEC optimized to act only on the consequential noise terms. This conclusion extends the analysis of Clifford circuits in \cite{filippov2024scalability}, which used the observable lightcone to discard some commuting error terms before applying PEC, but did not include the quadratic speedup available for PEC by discarding commuting terms within the lightcone.

The second sampling cost reduction comes from truncating terms in $\tilde{O}$, decreasing variance at a risk of biasing the result. This risk is reduced by an \textit{a priori} likelihood that most Paulis 
will individually have small expectation values, and will collectively accumulate net bias slowly via a random walk. Total truncation is simplest: instead of obtaining $\tilde{O}$ via an expensive classical computation, one simply approximates $\langle\tilde{O}\rangle \approx \langle O\rangle\prod_{t,i}(1-q_{t,i})$, functionally equivalent to representing all the noise as a single depolarizing channel. 
Mitigation then simplifies to a rescaling of the unmitigated $\langle O\rangle$, with a sampling cost of $\gamma_{\tilde{O}}^2 = \prod_{t,i}(1-q_{t,i})^2$. For small error rates, $\gamma_{\tilde{O}}^2\approx\gamma_{PEC}$, yielding a quadratic speedup, at a risk of truncation bias.

\section{Conjugating large Pauli-sum operators}
\label{app:zigzag}

This appendix describes an algorithm for approximating the conjugation of a Pauli-sum operator, $A$, by another such operator, $B$, by generating only the most significant terms in the product, $B A B^\dagger$ (see Algorithm~\ref{alg:heuristic-conjugation}).

Let
\[
  A = \sum_{s=0}^{N-1} \alpha_s P^A_s,
  \qquad
  B = \sum_{i=0}^{M-1} \beta_i P^B_i
\]
be Hermitian Pauli-sum operators, where each $P^A_s$ and $P^B_i$ is a Pauli operator and $\alpha_s,\beta_i \in \mathbb{R}$.
The exact conjugation,
\[
  A' = B A B^\dagger
    = \left( \sum_i \beta_i P^B_i \right)
      \left( \sum_s \alpha_s P^A_s \right)
      \left( \sum_j \beta_j P^B_j \right)
\]
requires evaluating all $M^2 N$ Pauli products $P^B_i P^A_s P^B_j$ and then merging duplicate Pauli terms, which becomes prohibitive for large operators. This likewise prohibits a naive approximate computation in which one first computes the exact result and then discards all but the desired number of largest terms. Instead, we require an approximate method in which only a limited number of Pauli products are computed.

We assume the Pauli-sum operators are stored as sorted lists:
\[
A = \{(\alpha_s, P^A_s)\}_{s=0}^{N-1},\qquad
B = \{(\beta_i, P^B_i)\}_{i=0}^{M-1},
\]
with $\lvert \alpha_{s+1} \rvert \le \lvert \alpha_s \rvert,$ $\lvert \beta_{i+1} \rvert \le \lvert \beta_i \rvert$.

We will use these sorted lists to generate terms in a priority max-queue, $Q$, holding index triplets, $(i,s,j)$, which represent a Pauli resulting from the multiplication of 3 Pauli terms, $P^B_iP^A_sP^B_j$. These index triplets are prioritized in the queue by the associated coefficient magnitude, $\lvert \beta_i \alpha_s \beta_j \rvert$.

To avoid evaluating all $M^2 N$ products, we search over the lattice of index triplets, $(i,s,j)$, in decreasing order of coefficient magnitude. We begin with the maximally weighted triplet, $(0,0,0)$, corresponding to $B_0 A_0 B_0$, then expand outward by pushing adjacent triplets, $(1,0,0)$, $(0,1,0)$, and $(0,0,1)$, into the priority queue. Each triplet popped from the queue yields one Pauli product contributing to the approximate $A'$. Terms are continually popped from the queue and their neighbors added to the queue until the maximum number of terms, $k$, have been popped from the queue and added to $A'$. This ordering ensures the $k$ kept terms are those with the largest coefficient magnitudes in the three-dimensional search space, the desired greedy criterion for better approximating $A'$. We note, however, that the Pauli products are not in general unique, and thus will not simply be the $k$ largest Pauli terms in $A'$ after de-duplication.

In cases where this search step limits overall performance, it can be significantly accelerated by setting the search step-size $\delta$ to be an integer greater than 1, or equivalently by searching only a coarse-grained sublattice of the three-dimensional space. Each kept point $(i,s,j)$ then indicates a cube of $\delta^3$ terms to include in $A'$, with corners at $(i+\{0,\delta\},s+\{0,\delta\},j+\{0,\delta\})$. This reduces the search time by a factor of $\delta^3$, at a cost of potentially deprioritizing some terms with larger coefficient magnitudes. A choice of $\delta=4$ made the search time relatively insignificant in our examples. Future implementations might further benefit from choosing the step size $\delta$ independently for each axis in the search space, perhaps inversely related to the gradient of the coefficient magnitudes in each operator.

\begin{algorithm}[H]
  \caption{Heuristic conjugation of Pauli operators}
  \label{alg:heuristic-conjugation}
  \begin{algorithmic}[1]
    \Require Sorted operator $A = \{(\alpha_s, P^A_s)\}_{s=0}^{N-1}$
    \Require Sorted operator $B = \{(\beta_i, P^B_i)\}_{i=0}^{M-1}$
    \Require Maximum number of terms to include, $k$
    \Ensure Approximate conjugated operator $A' \approx B A B^\dagger$ by including only the most significant terms
    \State Initialize empty Pauli operator, $A'$
    \State Initialize empty priority max-queue, $Q$
    \State Initialize empty visited set, $V$

    \State Insert $(0,0,0)$ into $Q$ with priority $\lvert \beta_0 \alpha_0 \beta_0 \rvert$
    \State Add $(0,0,0)$ to $V$
    \State $n_{\text{out}} \gets 0$ 
    \While{$Q \neq \emptyset \land n_{\text{out}} < k$}
      \State $(i,s,j) \gets \Call{PopMax}{Q}$
      \State $c' \gets \beta_i \alpha_s \beta_j$
      \State $P' \gets \Call{PauliProduct}{P^B_i P^A_s P^B_j}$
      \State $A'[P'] \gets A'[P'] + c'$ \Comment{accumulate and de-duplicate}
      \State $n_{\text{out}} \gets n_{\text{out}} + 1$
      \State $\mathcal{N} \gets \{(i{+}1,s,j),(i,s{+}1,j),(i,s,j{+}1)\}$ \Comment{neighbors}
      \ForAll{$(i',s',j') \in \mathcal{N}$}
        \If{$i' < M \land s' < N \land j' < M$}
          \If{$(i',s',j') \notin V$}
            \State $c_{\text{nb}} \gets \beta_{i'} \alpha_{s'} \beta_{j'}$
            \State Insert $(i',s',j')$ into $Q$ with priority $\lvert c_{\text{nb}} \rvert$
            \State Add $(i',s',j')$ to $V$
          \EndIf
        \EndIf
      \EndFor
    \EndWhile

    \State \Return $A'$
  \end{algorithmic}
\end{algorithm}

\section{Approximate application of maps $\{\Lambda_{t,i}^{\prime-1}\}$ in parallel}
\label{app:parallel}

The prospect of quantum-centric supercomputing with PNA requires parallelization of each classical step in the PNA algorithm, so as to meaningfully leverage the many cores and large amount of memory of an HPC (high-performance computing) system. The two major classical steps of the PNA computation are forward propagation of errors through unitary gates, and backwards propagation of the observable through the resulting antinoise maps.

The forward propagation, producing the set of forward-evolved Lindblad generators $\{E'_{t,i}\}$, is trivially parallelizable over all $E_{t,i}$ in the circuit, as the propagation of each unitary through the noiseless circuit is a fully independent problem. In principle the forward Pauli-propagation of a single unitary, i.e. the inner loop, may also be parallelized \cite{broers2025parallelprop} if enough classical resources are available to warrant nested parallelism, and if the communication overhead permits a benefit, though this was not done in our implementation for simplicity.

However, the backward propagation,
\begin{equation}
    \tilde{O} = \left(\underset{t,i}{\bigcirc}\Lambda'^{-1}_{t,i}\right) O.
\end{equation}
involves the serial application of each map to $O$, so cannot be exactly parallelized over maps $\Lambda_{t,i}^{\prime-1}$ since the input to each map requires first computing the output of the preceding map. Nonetheless, we can do a limited parallelization by means of an additional approximation, included as optional functionality in our implementation and discussed immediately below. We expect it is possible to parallelize the ``inner-loop'' accumulation of product terms $(i,s,j)$ within the application of each channel (App.~\ref{app:zigzag}), with no extra approximation necessary, but leave that to a future implementation.

For brevity, we combine the two indices $t,i$ into a single time index $T$ equal to the order in which the antinoise maps are processed, such that
\begin{equation}
    \tilde{O} = \left(\underset{T}{\bigcirc}\Lambda'^{-1}_{T}\right) O.
\end{equation}
To parallelize over maps over $C$ classical processing cores, we group the maps into batches of $C$ maps each:
\begin{equation}
    \tilde{O} = \left(\underset{b}{\overset{B}{\bigcirc}}\left(\underset{c=1}{\overset{C}{\bigcirc}}\Lambda'^{-1}_{T_c=c+bC}\right) \right)O.
\end{equation}
We focus on the result of applying a single batch to an observable,
\begin{equation}
    \tilde{O}_b = \left(\underset{c=1}{\overset{C}{\bigcirc}}\Lambda'^{-1}_{T_c}\right) O,
\end{equation}
and expand it in terms ordered by the number of errors $E_T$ acting on (conjugating) the observable,
\begin{equation}
\tilde{O}_b = \tilde{O}_b^{(0)} + \tilde{O}_b^{(1)} + \tilde{O}_b^{(2)} + ...
\end{equation}
where
\begin{align}
\tilde{O}_b^{(0)} &= O\prod_{c}^C(1-q_{T_c}) \label{eq:otilde_zero}\\
\tilde{O}_b^{(1)} &= \sum_{c}^C \bar{E}'_{T_c} \tilde{O}^{(0)} \bar{E}'_{T_c} \\
\tilde{O}_b^{(2)} &= \sum_{c_1<c_2}^C \bar{E}'_{T_{c_2}} \bar{E}'_{T_{c_1}} \tilde{O}^{(0)} \bar{E}'_{T_{c_1}} \bar{E}'_{T_{c_2}}.
\end{align}
The overbar represents absorption of quasiprobability into the operator as $\bar{E}_T = \sqrt{|q_T|/(1-q_T)}E_T$, and the apostrophe indicates evolution from the time the error occurs to the end of the circuit $E' = UEU^\dagger$, as in the main text.

The $C$ terms in $\tilde{O}_b^{(1)}$ may be computed in parallel, with each core running the conjugation routine of App.~\ref{app:zigzag}, providing the first-order approximation $\tilde{O}_b \approx \tilde{O}_b^{(0)}+\tilde{O}_b^{(1)}$. Note this means the required memory scales roughly with the number of cores in the simplest implementation. The merged, deduplicated results are then provided as the input for the next batch of maps. The approximation holds for batch size $C$ chosen small enough that $|q_TC|\ll1$ -- roughly, when the probability of more than one error occurring in any batch can be neglected. In the limiting case $C=1$ (no parallelization), the equality is exact. For superconducting hardware where we expect individual error-generator quasiprobabilities on scales like $\sim 10^{-4}$, we can leverage a finite number of classical processing cores $C \lesssim 10^2$. If desired, the batch size $C$ could be dynamically varied according to the known quasiprobabilities to maximize parallelization while keeping $|q_TC|$ small.

  Parallelization using this first-order approximation with fixed batch size is available in our open-source PNA implementation via an optional batch-size argument. (There, the number of cores may be different than the batch size, and cores dynamically switch between forward propagation to compute $\bar{E}'_T$ or backward propagation of $O$ through $\Lambda_T^{\prime-1}$ as needed).

  A more limited parallel computation of $\tilde{O}_b^{(2)}$ appears to be possible as well, which should enable a similarly accurate approximation while using moderately more cores. Restricting to the case where all error maps within a single batch commute, such as when each batch is contained within the Pauli noise model of a single layer with constant $t$, then after some algebra one can write
  \begin{equation}
      \tilde{O}_b^{(2)} = \frac{1}{2}\sum_{c}^C \bigg(\bar{E}'_{T_c} \tilde{O}_b^{(1)}  \bar{E}'_{T_c} - \left(\frac{q_{T_c}}{1-q_{T_c}}\right)^2 \tilde{O}_b^{(0)} \bigg).
  \end{equation}
  We suppose each individual core has previously computed $\bar{E}'_{T_c}$ and the results from each core have been merged to obtain $\tilde{O}_b^{(1)}$. Now, each core runs the routine from App.~\ref{app:zigzag} again to compute $\bar{E}'_{T_c} \tilde{O}_b^{(1)} \bar{E}'_{T_c}$, the contribution to $\tilde{O}_b^{(2)}$. It may be an interesting future research question to evaluate whether this extra step to compute the second-order contribution can be beneficial compared to simply reducing the batch size (and degree of parallelization) $C$ until the second-order contribution can be neglected.

\section{Invalidity of applying certain lightcone reductions before PNA}
\label{app:lightcones}

A standard procedure for simplifying quantum circuits is to eliminate all operations outside the lightcone of the measured observable, as these operations have no causal connection to the computed expectation value. A simple implementation deletes all gates outside the geometric lightcone of the observable, the region defined by starting from the observable and tracing left, up, or down along all connected wires and gates. A slightly more powerful, but still efficient, scheme works backwards from the observable, repeatedly checking if each gate encountered commutes with all subsequent operations including the observable, and removing the gate if so \cite{Kim2023utility, qiskit2024}. The procedure can be applied again with time reversed and with the state $\rho$ in place of the observable; since the expectation value is an overlap integral of $\rho$ and $O$, only operations in the overlap of their respective lightcones can participate. We broadly refer to all such techniques as lightcone reductions.

In the context of probabilistic error cancellation, lightcone reductions can remove not only gates but also noise channels from circuits, yielding sometimes-dramatic reductions in error-mitigation sampling costs \cite{tran2023locality, eddins2024lightcone}. A question, then, is to what extent lightcone methods can be used in combination with PNA, either to lower sampling-costs or to remove operations from the exponentially-difficult classical Pauli-propagation problems. A tempting approach is to apply the lightcone reduction first, and then to apply PNA. Perhaps surprisingly, this approach is not valid, leading to a modified observable $\tilde{O}$ that does not accurately cancel noise on a hardware experiment. The accuracy of PNA relies on pairing a modified observable with the noisy circuit performed by the hardware. However, the lightcone pass modifies this noisy circuit such that it no longer corresponds to the hardware execution. This can change the $\tilde{O}$ produced by PNA, giving an observable that accurately mitigates the lightcone-reduced noisy circuit, but not the noisy circuit actually run by the hardware.

\begin{figure}
    \includegraphics[width=0.66\linewidth]{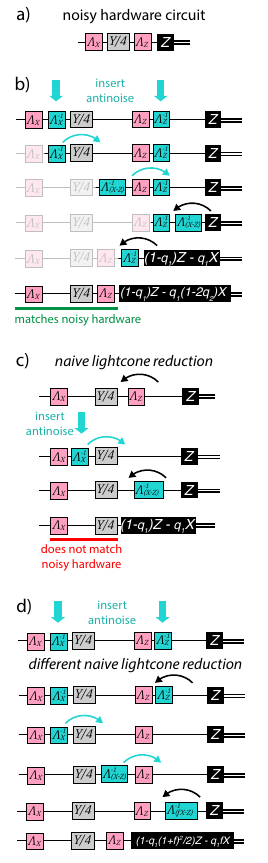}
    \caption{(a) A noisy one-qubit circuit we wish to mitigate, providing a very small cautionary example of how seemingly innocuous lightcone reductions lead to invalid results when naively combined with PNA. See text for definitions. (b) Standard PNA procedure, with curved arrows indicating forward propagation of the antinoise and backward propagation of the measured observable. (c) While it may look safe to delete the $\Lambda_Z$ just before the $Z$ measurement, this neglects the effect of $\Lambda_Z$ when measuring the $X$ component of $\tilde{O}$ on the noisy hardware, and thus fails to correctly compute $\tilde{O}$. (d) Keeping $\Lambda_Z$ but deleting $\Lambda_Z^{-1}$ also fails, which we attribute to the untracked correlation of the simulated and hardware copies of $\Lambda_Z$.}
    \label{fig:pna_lightcone}
\end{figure}

Figure~\ref{fig:pna_lightcone} works through three different attempts at applying PNA to an example noisy circuit with only one qubit (panel a). Here $\Lambda_X$ ($\Lambda_X^{-1}$) applies $X$ with probability $p_1$ (quasiprobability $q_1$), and $\Lambda_Z$ ($\Lambda_Z^{-1})$ applies $Z$ with probability $p_2$ (quasiprobability $q_2$). The standard PNA computation, with no lightcone reduction (Fig.~\ref{fig:pna_lightcone}b), is:
\begin{align}
    \tilde{O}&=\Lambda_Z^{-1}\left[\left(R_Y\left(\frac{\pi}{4}\right)\Lambda_X^{-1}R_Y\left(-\frac{\pi}{4}\right)\right)\left[Z\right]\right] \\
    &=\Lambda_Z^{-1}\left[\Lambda_{(X-Z)/\sqrt2}^{-1}\left[Z\right]\right] \\
    &=\Lambda_Z^{-1}\left[\left(1-q_1\right)Z+\frac{q_1}{2}(X-Z)Z(X-Z)\ \right] \\
    &=\Lambda_Z^{-1}\left[\left(1-q_1\right)Z-q_1X\right] \\
    &=(1-q_1)Z - q_1\Lambda_Z^{-1}\left[X\right] \\
    &=\left(1-q_1\right)Z-q_1\left(1-2q_2\right)X,
\end{align}
using square brackets to indicate the inputs to channels.

A naive lightcone reduction eliminates $\Lambda_Z^{-1}$(Fig.~\ref{fig:pna_lightcone}c), giving a different, incorrect result:
\begin{align}
\tilde{O}_{LC}&=\left(R_Y\left(\frac{\pi}{4}\right)\Lambda_X^{-1}R_Y\left(-\frac{\pi}{4}\right)\right)\left[Z\right]\\
&=\left(1-q_1\right)Z-q_1X.
\end{align}
This lightcone reduction is visibly invalid because the remaining noisy circuit does not match the noisy circuit we hope to mitigate -- it is missing one of the noise channels that the noisy hardware applies, and this noise channel cannot be trivially reintroduced.

This failure motivates asking, can we apply a lightcone reduction with a lighter touch, removing $\Lambda_Z^{-1}$ but leaving in $\Lambda_Z$ (Fig.~\ref{fig:pna_lightcone}d)? This turns out to fail as well, for a more subtle reason. Attempting to apply PNA, we evolve the single remaining antinoise map through the isolated noise channel $\Lambda_Z$, then apply the resulting antinoise map $\Lambda_Z\left[\Lambda_{(X-Z)/\sqrt{2}}^{-1}\right]$ to the measured observable Z:
\begin{align}
\tilde{O}_{LC2}&=\Lambda_Z\left[\Lambda_{(X-Z)/\sqrt{2}}^{-1}\right][Z] \\ 
&=\left(1-q_1\right)Z + \frac{q_1}{2}\Lambda_Z[X-Z]Z\Lambda_Z[X-Z].
\end{align}
The $Z$ noise channel acts on the $X$ component of the unitary error with a fidelity $f = (1-2p_2) \leq 1$:
\begin{align}
    \Lambda_Z\left[X-Z\right]&=\left(1-p_2\right)(X-Z)+p_2Z(X-Z)Z\\
    &=\left(1-2p_2\right)X-Z\\
    &=fX-Z.
\end{align}
Plugging this in gives
\begin{align}
    \tilde{O}_{LC2} &=(1-q_1)Z + \frac{q_1}{2}(fX-Z)Z(fX-Z) \\
    &=(1-q_1)Z + \frac{q_1}{2}((1-f^2)Z-2fX) \\
    &=\left(1-\frac{q_1}{2}(1+f^2)\right)Z - q_1fX
\end{align}
This answer is incorrect, $\tilde{O}_{LC2}\neq\tilde{O}$. We attribute the logical fault to classically evolving the map $\Lambda_{(X-Z)/\sqrt{2}}^{-1}$ through $\Lambda_Z$ without taking into account correlations between the action of $\Lambda_Z$ on $\Lambda_{(X-Z)/\sqrt{2}}^{-1}$, and its action on the quantum state $\rho$ during hardware execution. By treating these effects as independent, we include nonphysical trajectories in which $\Lambda_Z$ acts on the channel $\Lambda_{(X-Z)/\sqrt{2}}^{-1}$ with a $Z$ error, but nonsensically does not apply the same $Z$ error to the quantum state. A similar pitfall of treating two references to the same noise source as two independent noise sources is also considered in an appendix of \cite{eddins2024lightcone}. This issue is also why PNA processes errors in the circuit with an ordering that avoids evolving one map through another.

We conclude from these examples that the most straightforward applications of lightcone reductions to PNA are not valid. We have not ruled out that some form of lightcone reduction may yet prove beneficial. Means of combining the “lightcone shading” technique \cite{eddins2024lightcone} with PNA would be particularly interesting, as the two classical computations are so similar that their merger may incur minimal classical computational cost beyond running either alone.

\begin{figure*}[t]
\centering
\includegraphics[width=0.98\textwidth]{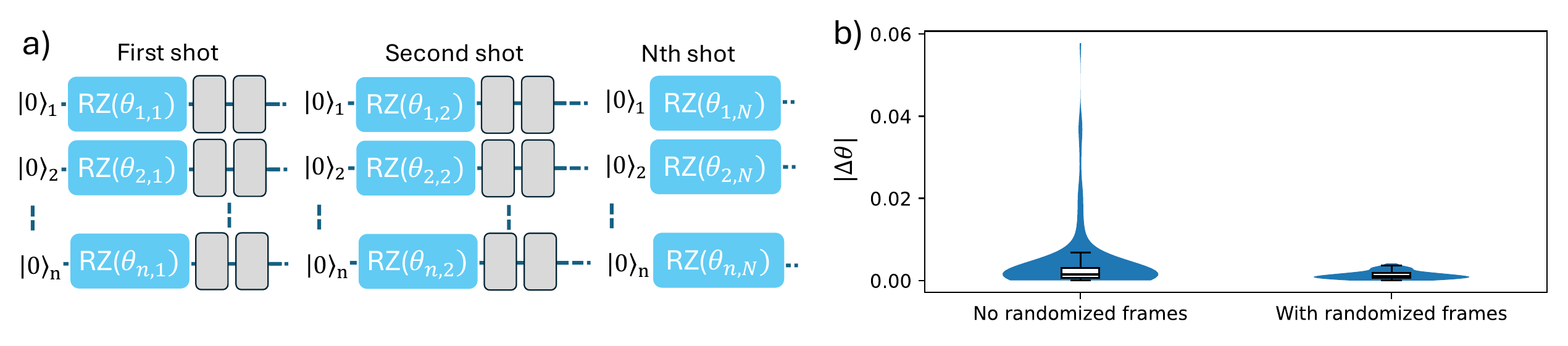}
\caption{
(a) Schematic of the randomized frame method for suppressing coherent drive crosstalk.
At the beginning of each circuit execution, a virtual $R_Z$ rotation with a randomly chosen angle is applied to every qubit. In the absence of crosstalk, these rotations are equivalent to the identity and do not affect the ideal circuit. However, in the presence of microwave drive crosstalk, the random $R_Z$ rotations change the effective rotation axis of the crosstalk Hamiltonian from shot to shot, leading to suppression of coherent errors on average. (b) Distribution of $|\Delta\theta_i|=|\theta_i^{S}-\theta_i^{B}|$ for $X$-gate error amplification on an IBM Heron device using 146 qubits. $\theta_i^{S}$ is over‑rotation angle estimated when characterizing all qubits simultaneously and $\theta_i^{B}$ is when characterizing in smaller batches. Randomized frames narrow the distribution, consistent with suppression of coherent crosstalk.
}
\label{fig:random_frames}
\end{figure*}

\section{Device error distributions}
\begin{figure}[H]
    \centering
  \includegraphics[width=.9\linewidth]{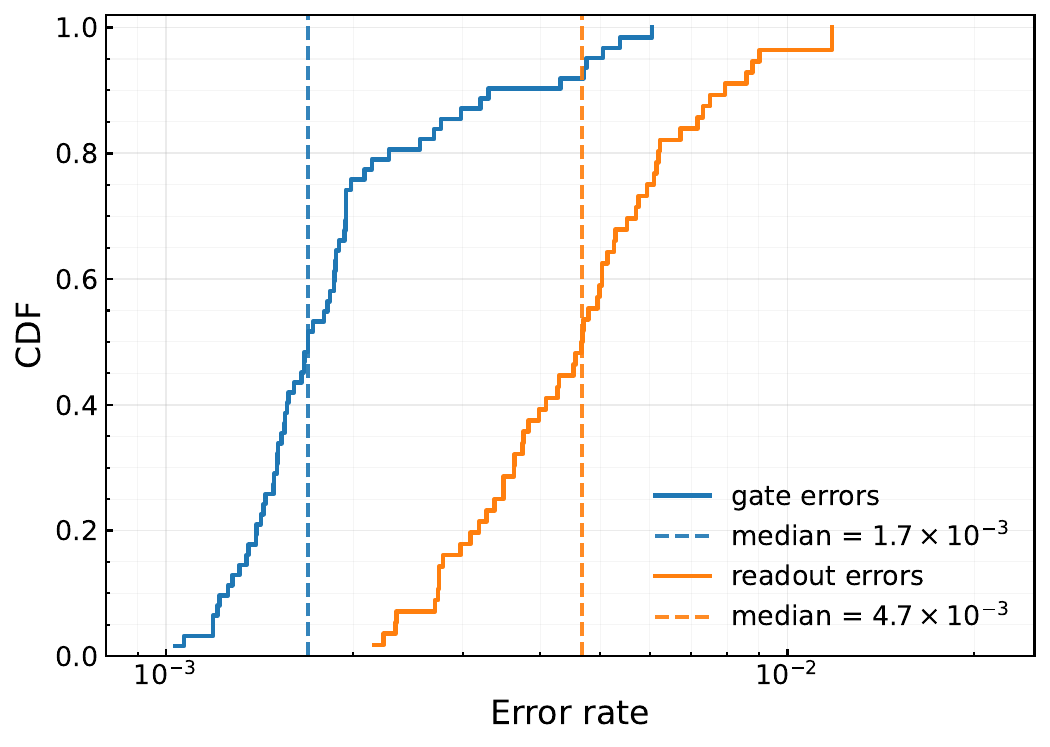}
  \caption{Cumulative distribution functions of two-qubit error rates (blue) and readout error rates (orange) reported by \texttt{ibm\_boston} near the time of the main-text experiments. The rates shown correspond to the 56-qubit layout (Fig.~\ref{fig:pna_expt}b inset).}
  \label{fig:CDFs}
\end{figure}

\section{Mitigating drive-crosstalk errors using random frame updates}\label{app:random_frames}

In superconducting quantum processors, microwave pulses intended to implement gates on a target qubit may also drive other qubits due to classical crosstalk in the microwave control lines, which can often be nonlocal with respect to the qubits' connectivity~\cite{nuerbolati2022canceling,PRXQuantum.5.030350,wesdorp2026mitigatingcrosstalkerrorssimultaneous}. Here we describe a procedure to suppress bias due to this effect, which were used in calibrations for our experiments.

As a minimal model, we consider two qubits where we drive qubit 1 at its frequency $\omega_1$. Applying pulse envelope $\Omega_1(t)$ on qubit 1 drives qubit 2 via classical crosstalk with a similar pulse shape and with a relative amplitude and phase characterized by the dimensionless coefficient $c_{1\to2}$ and phase offset $\phi_{1\to2}$. 
In the rotating frame of the qubits and after applying the rotating-wave approximation, the resulting crosstalk Hamiltonian acting on qubit 2 is

\begin{equation}
\begin{split}
    H(t)/\hbar = c_{1\to2} & \frac{\Omega_1(t)}{2} [ X_2\cos(\Delta_{21} t + \phi_2 - \phi_1 + \phi_{1\to2}) \\
    & + Y_2\sin(\Delta_{21} t + \phi_2 - \phi_1 + \phi_{1\to2})]
\end{split}
\end{equation}
with $\Delta_{21}=\omega_2-\omega_1$, and $\phi_1$ and $\phi_2$ denote the drive phases of each qubit respectively. When the qubit frequencies are nearly resonant, such crosstalk can lead to direct excitation; otherwise, it can be considered non-Markovian off-resonant error \cite{PhysRevApplied.21.024018}.
This interaction is especially harmful in error amplification circuits for gate calibration, where gates are repeatedly applied with fixed relative phases, as coherent crosstalk can accumulate constructively over the circuit depth, leading to systematic errors and biased estimates of gate parameters.

Here we introduce a simple and hardware-efficient mitigation strategy that suppresses such coherent crosstalk when averaging over multiple experimental shots. As illustrated in Fig.~\ref{fig:random_frames}(a), at each circuit initialization we apply a random virtual $R_Z$ rotation to each qubit. This randomizes the relative phase between qubits from shot to shot, thereby changing the effective rotation axis associated with the crosstalk Hamiltonian. This keeps the exact same intended circuit, while coherent crosstalk contributions average out. 

Results of applying this method to error amplification of $X$ gates are shown in Fig.~\ref{fig:random_frames}(b). We measure the over‑rotation of the $X$ gates on an IBM Heron device device, \texttt{ ibm\_kingston},  using 146 qubits.
For each qubit $i$, we compare the rotation angle measured when all qubits operate simultaneously, $\theta_i^{S}$, with the reference value $\theta_i^{B}$ obtained by partitioning the device into smaller groups consist of 16 batches of distant qubits and running each batch separately. The distribution of $|\Delta\theta_i|=|\theta_i^{S}-\theta_i^{B}|$ is significantly narrowed when randomized frames are applied, indicating suppression of coherent crosstalk, therefore enabling more accurate per‑qubit gate calibration.

\onecolumngrid
\section{Details on Euclid theory}

\label{sec:euclid_theory_app}

In this section, we provide more details and examples on the Euclid perturbative series expansion for the map $\mathcal{D} =(\mathcal{C}_\text{ideal}\ \mathcal{C}_\text{noisy}^{-1})^\dagger$ applied to the target observable. \\

Although backpropagation of $O$ through the original circuit is generally prohibitively difficult, propagation through $\mathcal{D}$ can be easier due to the internal structure of the map. In the spirit of Clifford Perturbation Theory (CPT)~\cite{begusic2025cpt, begusic2024fast}, we expand $\tilde{O} = \mathcal{D}O$ as a sum over single-Pauli trajectories backwards through $\mathcal{D}$. We make use of the following unitary and noisy propagation rules (see also Eqs. \eqref{eq:evolve_gate}, \eqref{eq:evolve_noise} in the main text):
\begin{equation}
    U_t P U_t^\dagger = \left(\cos(\theta_t) -i\sin(\theta_t)R_t\right)^{\langle P, R_t \rangle} P
\end{equation}
and
\begin{equation}
    U_t^\dagger P U_t = \left(\cos(\theta_t) + i\sin(\theta_t)R_t\right)^{\langle P, R_t \rangle} P
\end{equation}
for the evolution of a Pauli $P$ past a single gate $U_t = \exp(-iR_t (\theta_t/2))$, as well as \begin{equation}\label{eq:evolve_noise_fid}
    \Lambda_t^{-1}(P)= P/f_t^{P} \approx (1+\epsilon_t^P)P, 
\end{equation}
for inverse noise, where 
$f_t^P=1-\epsilon_t^P$ is the Pauli fidelity of $P$ acted on by the noise map $\Lambda_t$. The approximation in the last Equation holds in the weak noise regime, and the symplectic inner product $\langle P, R_t \rangle$ is 0 (1) if $P$, $R_t$ (anti-)commute. In this way, one immediately writes
\begin{equation}
    \tilde{O} \approx
    \sum_{k=0}^{\tilde{N}}\ \sum_{1 \leq j_1 < \cdots < j_k \leq N_\text{g}} c_{j_1\ldots j_k} P_{j_k} \cdots P_{j_1}O,
\label{eq:CPT-EM-general}
\end{equation}
where the complex numbers $c_{j_1\ldots j_k}$ contain combinations of trigonometric factors and Pauli fidelities, $N_\text{g}$ is the number of unitary gates in $\mathcal{C}_{\text{ideal}}$, and $\tilde{N}$ limits the order of the expansion in the number of Pauli rotation generators applied to $O$. \\

We classify terms in the expansion of $\mathcal{D}$ using diagrams (see Fig.~\ref{fig:Euclid-Diagrams}) which help to collect all terms in Eq.~\eqref{eq:CPT-EM-general} containing up to a maximum number of sines of gate angles, which can be made small by expressing the circuit in the Clifford interaction picture. In a diagram, each continuous trajectory represents one Pauli path in the backpropagation of $O$ through $\mathcal{D}$. The horizontal dimension is time, and each node indicates a noiseless ($\mathcal{U}_j$) or inverse noisy ($\Lambda_j^{-1} \mathcal{U}_j^{-1}$) operation. The vertical dimension represents the discrete set of all multiqubit Paulis, with steps indicating transformations from one Pauli to another.

We draw a \textit{step} when a branching point at position $j$, either in $\mathcal{C}_\text{ideal}$ or $\mathcal{C}_\text{noisy}^{-1}$, takes a $\sin\theta$ path, modifying the observable to $P_jO$ left of the step. A diagram that terminates $k$ steps away from $O$ will contribute to the coefficient of a term where $k$ Pauli generators act on $O$. We call $k$ the \textit{Pauli order} of a term. A term with Pauli order $k$ will have at least $k$ factors of $\sin\theta$.

\textit{Bubbles}, instead, consist of a pair of symmetric steps in $\mathcal{C}_\text{ideal}$ and $\mathcal{C}_\text{noisy}^{-1}$, centered in the diagram, 
which do not modify the terminal Pauli of the diagram, (i.e., they do not affect the Pauli order). However, they temporarily change the Pauli being backpropagated, which can in turn change whether it commutes with a gate or noise channel. Ultimately, bubble diagrams yield additive contributions to the coefficient $c_{j_1\ldots j_k}$ of the relevant $P_{j_k} \cdots P_{j_1}O$ term. 
A diagram with $p$ bubbles yields a coefficient with at least $p$ factors of $\sin^2\theta$; we call $p$ the \textit{Dyson order}. The generalization to combinations of bubbles and steps is relatively straightforward (see the next subsections). \\

The noise-canceling observable $\tilde{O}$ of Eq.~\eqref{eq:CPT-EM-general} can be iteratively grown in both Pauli and Dyson orders: the former controls the number of independent 
$P_{j_k} \cdots P_{j_1}O$
terms, while the latter controls the accuracy of the coefficients $c_{j_1\ldots j_k}$. Concretely, we write
\begin{equation}
  c_{j_1\ldots j_k} = \sum_{p=0}^{\tilde{P}} b_p^{\{j_1\ldots j_k\}}
  \label{eq:CPT-em-coefficients-general}
\end{equation}
where $\tilde{P} \leq N-k$ is the maximum Dyson order allowed in the expansion. Each individual contribution $b_p^{\{j_1\ldots j_k\}}$ with Pauli order $k$ and Dyson order $p$ corresponds to one or more diagrams with $k$ steps and $p$ bubbles, and contains $\kappa = k+2p$ factors of $\sin\theta$. We refer to $\kappa$ as the \textit{CPT order}. \\

As in CPT, we will assume that the powers of $\sin\theta$ can be used to rank the terms appearing in the error mitigation series, with higher powers corresponding to smaller contributions. 
Crucially, a well-formed, consistent treatment up to, say, order $\sin^K\theta$, needs to take into account all terms with CPT order $\kappa \leq K$. This means that the coefficients for terms $P_{j_k} \cdots P_{j_1}O$ with lower Pauli order will be evaluated by including diagrams of higher Dyson order (with more bubbles). Conversely, only pure step diagrams are allowed for $k=K$, and the approximation series terminates. In the following, we provide some explicit examples of analytical diagram contributions and of the subsequent perturbative expansion.

\subsection{Diagrammatic notation}

\begin{figure}[htbp!]
  \centering
  \includegraphics[width=0.5\columnwidth]{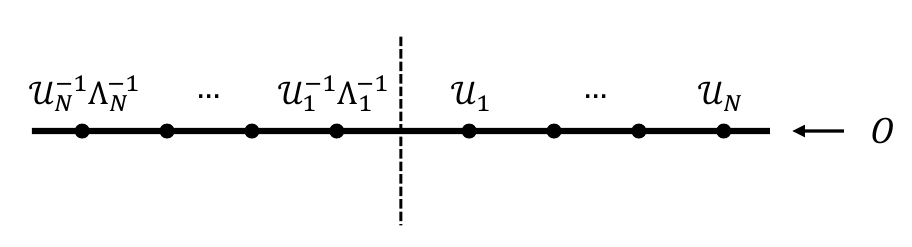}
  \caption{Graphical representation of a \emph{line diagram}. The target operator $O$ is backpropagated from the right to the left. One first applies the inverse noisy map (first half of the diagram), and then the noiseless map (second half of the diagram).}
  \label{fig:LineDiagram}
\end{figure}

We begin by introducing the \textit{line} diagram (see Figure~\ref{fig:LineDiagram}).
This schematically represents the direct (on the right) and noisy inverse (on the left) parts of $\mathcal{D}$, with points on the line (also called \textit{nodes} from now on) denoting operations of the form $\mathcal{U}_i$ and $\mathcal{U}^{-1}_i\Lambda^{-1}_i$ respectively. 
To keep notation simple, below we will not show the discrete points explicitly.
Computing the expansion of $\mathcal{D}$ is equivalent to propagating the target observable $O$ from right to left in the diagram, applying at each node the rules introduced in Eq.~\eqref{eq:evolve_gate}-\eqref{eq:evolve_noise}. In the following, we will define as \emph{backpropagated operator} at a given qubit diagram node the operator that is obtained by applying all layers up to that node.

At each node of the line diagram, $O$ is left unchanged either because it commutes with the relevant gate generator or because the $\cos\theta$ path is taken. However, the observable picks up a product of coefficients $(1+\epsilon_i^{O})$ from the application of the inverse noise maps. The resulting contribution then reads
\begin{equation}
  \mathbb{L}[O] = \Big[\prod_{\substack{j=1 \\ \{P_j,O\}}}^N \cos^2\theta_j \Big]
                  \Big[\prod_{j=1}^N (1+\epsilon_j^O)\Big]\cdot [O]\, ,
  \label{eq:line-term}
\end{equation}
where in the first product, $\{P_j,O\}$ is a short hand notation to indicate that a $\cos^2\theta_j$ term only appears if $\{P_j,O\}=0$. Notice that each cosine appears squared since both halves of the diagram contribute equally. \\

\begin{figure*}
    \centering
    \includegraphics[width=\textwidth]{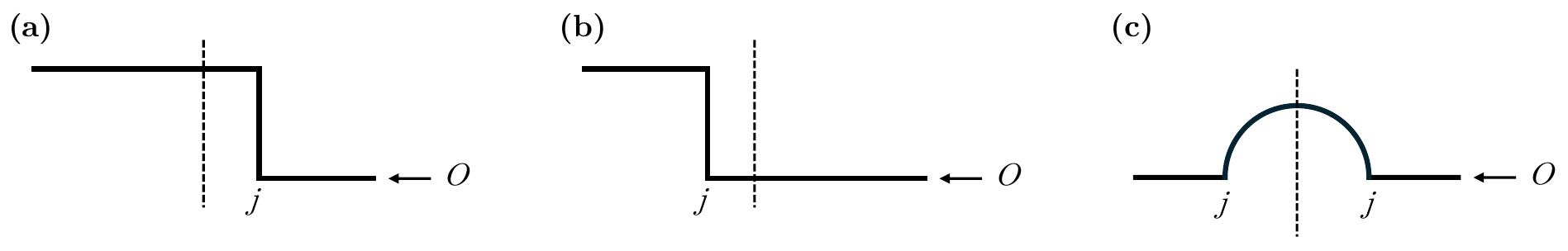}
    \caption{Examples of step and bubble diagrams: (a) direct 1-step diagram $\mathcal{Z}_j [O]$, (b) inverse 1-step diagram $\mathcal{Z}'_j [O]$, (c) a single bubble forming on top of the line diagram, i.e., $\Omega_j [O]$.}
    \label{fig:steps-omega-intro}
\end{figure*}

\textbf{Step diagrams.} When $O$, at single node in either in the direct or inverse part of $\mathcal{D}$, follows the $\sin\theta$ path, the 1-\textit{step} diagram is generated. In this case, if the step occurs at a position $j$, the observable is modified to $P_jO$ and all noise factors collected after the step should be adjusted accordingly. The possible resulting contributions then take the form
\begin{equation}
    \mathcal{Z}_j [O]
        = i\delta_{\{P_j,O\}} \cos\theta_j \sin\theta_j 
          \Big[\prod_{\substack{k=j+1 \\ \{P_k,O\}}}^N \cos\theta_k\Big]
          \Big[\prod_{\substack{k=j+1 \\ \{P_k,P_jO\}}}^N \cos\theta_k\Big] 
          \Big[\prod_{\substack{k=1 \\ \{P_k,P_jO\}}}^{j-1} \cos^2\theta_k\Big]
          \Big[\prod_{k=1}^N (1+\epsilon_k^{P_jO})\Big]\cdot[P_jO]
  \label{eq:Z-term}
\end{equation}
if the step occurs in the direct part of the diagram (see Fig.~\ref{fig:steps-omega-intro}a), or
\begin{equation}
  \mathcal{Z}'_j [O]
        = -i\delta_{\{P_j,O\}} \cos\theta_j \sin\theta_j 
            \Big[\prod_{\substack{k=j+1 \\ \{P_k,O\}}}^N \cos\theta_k\Big]
            \Big[\prod_{\substack{k=j+1 \\ \{P_k,P_jO\}}}^N \cos\theta_k\Big] 
            \Big[\prod_{\substack{k=1 \\ \{P_k,O\}}}^{j-1} \cos^2\theta_k\Big]
            \Big[\prod_{k=j+1}^N (1+\epsilon_k^{P_jO})\Big]
            \Big[\prod_{k=1}^{j} (1+\epsilon_k^{O})\Big] \cdot [P_jO]
  \label{eq:Zprime-term}
\end{equation}
for a step in the inverse part (see Fig.~\ref{fig:steps-omega-intro}b). In the above expressions, $\delta_{\{P_j,O\}}$ is a shorthand notation for the constraint $\{P_j, O\} = 0$ that must be satisfied for the diagram to appear in the Pauli Propagation-like expansion. Notice that it is no longer the case that all $\cos\theta$ contributions appear in pairs, since some of the constraints on the products are different, and one $\sin\theta$ is included. Step diagrams can naturally be generalized to multiple steps (see below).

\textbf{Bubble diagrams.} The last class of diagrams that may occur includes those cases in which one or more nodes points follow the $\sin\theta$ path at corresponding positions in both the direct and inverse part of the map. These \textit{bubbles} can only form in a symmetric fashion across the middle point of the diagram, and -- due to the fact that Pauli operators square to the identity -- they do not modify the final observable to which the diagram itself leads (i.e., they do not influence the Pauli order). However, they temporarily change how the noise factors are picked up, as well as some of the anti-commutation checks, and ultimately yield an additive contribution to the coefficient $c_{j_1\ldots j_k}$ of the relevant $P_{j_k} \cdots P_{j_1}O$ term. As an example, consider a single bubble forming on top of the line diagram (Fig.~\ref{fig:steps-omega-intro}c):
\begin{equation}
    \Omega_j [O] = \delta_{\{P_j,O\}} \sin^2\theta_j \Big[\prod_{\substack{k=j+1 \\ \{P_k,O\}}}^N \cos^2\theta_k\Big]\Big[\prod_{k=j+1}^N (1+\epsilon_k^O)\Big]\Big[\prod_{\substack{k=1 \\ \{P_k,P_jO\}}}^{j-1} \cos^2\theta_k\Big]\Big[\prod_{k=1}^{j} (1+\epsilon_k^{P_jO})\Big] \cdot [O]
    \label{eq:bubble-term}
\end{equation}
As one can immediately see, the resulting observable is unchanged with respect to the baseline diagram ($O$ in this case), and the coefficient contains a $\sin^2\theta$ factor. The generalization to multiple bubbles and to combinations of step and bubbles is relatively straightforward. Notice that, in principle, if a bubble occurs across a step one may need to commute some of the Pauli operators applied to $O$ before using the identity $P^2 = \mathbb{I}$ -- however, this produces at most an additional sign.

\subsection{Example: perturbation theory for the $O$ branch}

\begin{figure*}
    \centering
    \includegraphics[width=\textwidth]{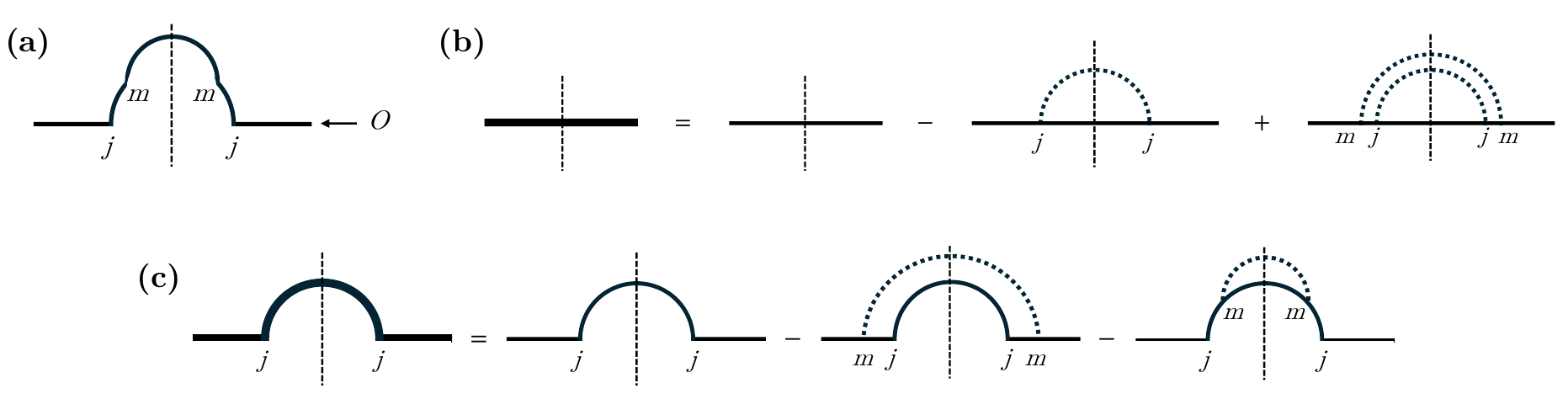}
    \caption{Contributions to $c_\varnothing$ up to order $\kappa=4$. (a) 2-bubble diagram $\Omega^{(2)}_{jm}[O]$, (b) decomposition of the line diagram into a thin line and virtual 1- and 2-bubble contributions, (c) decomposition of a 1-bubble diagram into its thin version and terms with one real plus one virtual bubble.}
    \label{fig:o-branch}
\end{figure*}

As a warm-up example, let us consider the construction of $c_{\varnothing}$, namely the coefficient of $O$ (i.e., the $k=0$ component) in Eq.~\eqref{eq:CPT-EM-general} up to a maximum CPT order $\kappa = 4$. The contributing diagrams are: the line term (Eq.~\ref{eq:line-term}, with $k=0$ and $p=0$), all the 1-bubble terms (Eq.~\ref{eq:bubble-term}, with $k=0$ and $p=1$) and all the 2-bubble terms (with $k=0$ and $p=2$). The latter are of the form (see Fig.~\ref{fig:o-branch}a)
\begin{equation}
\begin{split}
     \substack{\Omega \\ \Omega}_{jm}[O] \equiv \Omega^{(2)}_{jm}[O] = \, & \delta_{\{P_j,O\}}\delta_{\{P_m,P_jO\}} \sin^2\theta_j\sin^2\theta_m \Big[\prod_{\substack{l=j+1 \\ \{P_l,O\}}}^N\cos^2\theta_l \prod_{l=j+1}^N(1+\epsilon_l^{O})\Big] \\
     & \Big[\prod_{\substack{l=m+1 \\ \{P_l,P_jO\}}}^{j-1}\cos^2\theta_l \prod_{l=m+1}^{j}(1+\epsilon_l^{P_jO})\Big]\Big[\prod_{\substack{l=1 \\ \{P_l,P_mP_jO\}}}^{m-1}\cos^2\theta_l \prod_{l=1}^{m}(1+\epsilon_l^{P_mP_jO})\Big]\cdot [O]
\end{split}
\end{equation}
for $j=1,\ldots,N$ and $m=1,\ldots,j-1$. With some abuse of notation, the contributions to $c_{\varnothing}$ therefore read
\begin{equation}
    c_{\varnothing} \simeq b_0^{\{\varnothing\}} + \sum_{j=1}^N \Omega_j + \sum_{j=1}^N\sum_{m=1}^{j-1}  \Omega_{jm}^{(2)} \,
\end{equation}
where $b_0^{\{\varnothing\}}$ is the coefficient on the r.h.s.\ of Eq.~\eqref{eq:line-term} and the other quantities are also understood as the coefficients multiplying $O$ in the corresponding expressions. To proceed with the perturbative calculation, we will now make use of the well-known trigonometric identity $\cos^2\theta = 1 - \sin^2\theta$. With this, we will be able to rewrite all coefficients solely in terms of perturbative $\sin\theta$ parameters. Furthermore, we will also approximate the noise terms to first order as, for instance,
\begin{equation}
    \prod_{j} (1+\epsilon_j^O) \simeq 1 + \sum_j \epsilon_j^O\,.
\end{equation}
For the line diagram up to order $\sin^4\theta$ we can then write
\begin{equation}
     b_0^{\{\varnothing\}} \simeq \Big[1+\sum_{j=1}^N\epsilon_j^O\Big]\Big[\prod_{\substack{j=1 \\ \{P_j,O\}}}^N (1-\sin^2\theta_j) \Big] = \Big[1+\sum_{j=1}^N\epsilon_j^O\Big]\Big[1-\sum_{\substack{j=1 \\ \{P_j,O\}}}^N\sin^2\theta_j\big[1-\sum_{\substack{m=j+1 \\ \{P_m,O\}}}^N\sin^2\theta_m[1-\ldots]\big]\Big]
\end{equation}
Interestingly, we can interpret the r.h.s.\ of this equation as follows (see also Fig.~\ref{fig:o-branch}b): the first term, 
\begin{equation}
    [1+\sum_{j=1}^N\epsilon_j^O] \cdot 1    
\end{equation}
is a \textit{thin} line diagram, contributing noise factors at the zeroth perturbative order in $\sin\theta$. The second term
\begin{equation}
     - [1+\sum_{j=1}^N\epsilon_j^O] \cdot \sum_{\substack{j=1 \\ \{P_j,O\}}}^N\sin^2\theta_j
\end{equation}
is a \textit{virtual} 1-bubble diagram. Contrary to standard bubble diagrams, the noise factors are determined by the \textit{parent} line diagram. However, the trigonometric factors are the same of real bubble diagrams. As we will see below, virtual diagrams spawned by a real one at a given Dyson order essentially re-normalize the contributions of the corresponding real ones at higher Dyson orders. Finally, the third term 
\begin{equation}
     [1+\sum_{j=1}^N\epsilon_j^O] \cdot \sum_{\substack{j=1 \\ \{P_j,O\}}}^N \sum_{\substack{m=j+1 \\ \{P_m,O\}}}^N \sin^2\theta_j \sin^2\theta_m
\end{equation}
is a \textit{virtual} 2-bubble diagram. Notice that virtual bubbles are independent from each other, in the sense that the presence of a larger bubble does not modify the anti-commutation constraints of smaller ones. In pictures, this will be made explicit by the fact that virtual bubbles do not lay on each other.

Following the same steps for the real 1-bubble diagrams we get
\begin{equation}
    \Omega_j \simeq \delta_{\{P_j,O\}} \Big[1+\sum_{l=1}^j \epsilon_l^{P_jO} + \sum_{l=j+1}^N \epsilon_l^O\Big]\sin^2\theta_j 
    \Big[1-\sum_{\substack{m={j+1} \\ \{P_m,O\}}}^N\sin^2\theta_m[1-\ldots]\Big]
    \Big[1-\sum_{\substack{m={1} \\ \{P_m,P_jO\}}}^{j-1}\sin^2\theta_m[1-\ldots]\Big]
\end{equation}
which, up to order $\sin^4\theta$, yields a thin 1-bubble diagram and two diagrams with one real and one virtual bubble (see Fig.~\ref{fig:o-branch}c). Finally, for the real 2-bubble contributions we have
\begin{equation}
\begin{split}
     \Omega^{(2)}_{jm} \simeq \delta_{\{P_j,O\}}\delta_{\{P_m,P_jO\}} \sin^2\theta_j\sin^2\theta_m \Big[1+\sum_{l=j+1}^N\epsilon_l^{O}+\sum_{l=m+1}^{j}\epsilon_l^{P_jO}+\sum_{l=1}^{m}\epsilon_l^{P_mP_jO}\Big] \Big[1-\ldots \Big]
     \Big[1-\ldots\Big]\Big[1-\ldots\Big]\, ,
\end{split}
\end{equation}
i.e., only the corresponding thin diagram. Putting everything together we get, after some simple algebraic manipulations, a re-summed series of the form
\begin{equation}
    \begin{split}
    c_{\varnothing} \simeq \, & 1+\sum_{l=1}^N\epsilon_l^O
        + \sum_{\substack{j=1}}^N \sin^2\theta_j  \Bigg[\delta_{\{P_j,O\}}\sum_{l=1}^j (\epsilon_l^{P_jO} - \epsilon_l^O) + \\
        + & \sum_{\substack{m=j+1}}^N\delta_{\{P_m,O\}} \sin^2\theta_m \Big[\sum_{l=1}^{j}\delta_{\{P_j,P_mO\}}(\epsilon_l^{P_jP_mO} - \epsilon_l^{P_mO}) + \delta_{\{P_j,O\}}(-\epsilon_l^{P_jO} + \epsilon_l^O)\Big]\Bigg]
    \end{split}
\end{equation}
One may now observe that the zero-order contribution is just an effective depolarizing-like limit in which the noisy expectation value of the target observable is rescaled by the fidelity $1+\sum_{j=1}^N\epsilon_j^O$, independently from any coherent rotation angle. Higher order contributions depend instead on trigonometric factors, anti-commutation checks and differences of noise factors, where the latter arise from the re-normalization effects induced by virtual bubbles. 

Once again, let us stress that knowing $c_{\varnothing}$ at order $\sin^4\theta$ is not enough to obtain a fully consistent perturbative CPT expansion at order $\kappa = 4$. In fact, one would need to compute $c_{j_1}$ and $c_{j_1j_2}$ up to Dyson order $p=1$, as well as $c_{j_1j_2j_3}$ and $c_{j_1j_2j_3j_4}$ at Dyson order $p=0$. We will outline the first of these calculations in the next subsection.

\subsection{Perturbation theory of step diagrams: the $P_j O$ branch}

\begin{figure*}
    \centering
    \includegraphics[width=\textwidth]{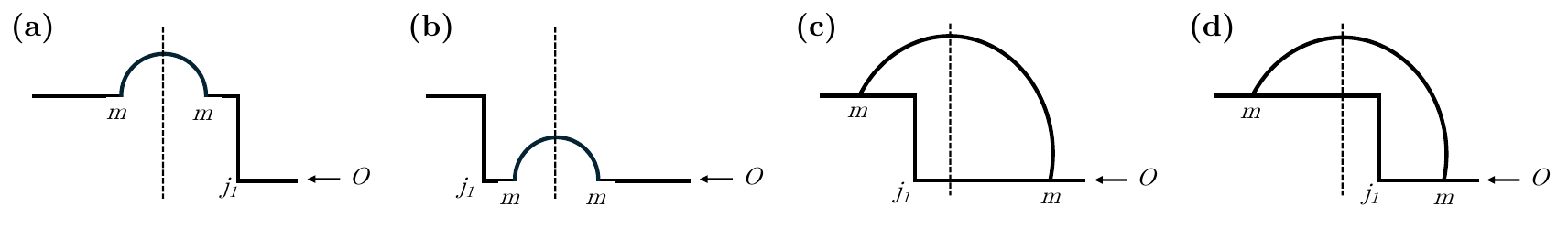}
    \caption{Contributions to the $P_{j_1}O$ branch with one step and one bubble. (a) $\Omega_m[\mathcal{Z}_{j_1}[O]]$, (b) $\mathcal{Z}'_{j_1}[\Omega_m[O]]$, (c) $\substack{\Omega_m \\ \mathcal{Z}_{j_1} }[O]$, (d) $\substack{\Omega_m \\ \mathcal{Z}'_{j_1} }[O]$.}
    \label{fig:pjo-branch}
\end{figure*}

Let us now consider the construction of one of the coefficients at Pauli order $k=1$, namely $c_{j_1}$, up to CPT order $\kappa=4$. The two base diagrams leading to $P_{j_1}O$ are the steps $\mathcal{Z}_{j_1} [O]$ and $\mathcal{Z}'_{j_1} [O]$ introduced in Eqs.~\eqref{eq:Z-term}-\eqref{eq:Zprime-term}. In addition to these, we need to consider the case of a single bubble forming on the $P_{j_1}O$ leg of $\mathcal{Z}_{j_1}$, namely (see Fig.~\ref{fig:pjo-branch}a)
\begin{equation}
\begin{split}
    & \Omega_m[\mathcal{Z}_{j_1}[O]] =  i\delta_{\{P_{j_1},O\}} \cos\theta_{j_1} \sin\theta_{j_1} \Big[\prod_{\substack{l={j_1}+1 \\ \{P_l,O\}}}^N \cos\theta_l\Big]\Big[\prod_{\substack{l={j_1}+1 \\ \{P_l,P_{j_1}O\}}}^N \cos\theta_l\Big] \Big[\prod_{l={j_1}+1}^N (1+\epsilon_l^{P_{j_1}O})\Big] \\
    \cdot &\, \delta_{\{P_m,P_{j_1}O\}} \sin^2\theta_m \Big[\prod_{\substack{l=m+1 \\ \{P_l,P_{j_1}O\}}}^{{j_1}-1} \cos^2\theta_l\Big]\Big[\prod_{l=m+1}^{{j_1}} (1+\epsilon_l^{P_{j_1}O})\Big]\Big[\prod_{\substack{l=1 \\ \{P_l,P_mP_{j_1}O\}}}^{m-1} \cos^2\theta_l\Big]\Big[\prod_{l=1}^{m} (1+\epsilon_l^{P_mP_{j_1}O})\Big] \cdot [P_{j_1}O]
\end{split}
\end{equation}
as well as the 1-bubble diagram on the $O$ leg of $\mathcal{Z}'_{j_1}$ (see Fig.~\ref{fig:pjo-branch}b)
\begin{equation}
\begin{split}
    & \mathcal{Z}'_{j_1}[\Omega_m[O]] =  -i\delta_{\{P_{j_1},O\}} \cos\theta_{j_1} \sin\theta_{j_1} \Big[\prod_{\substack{l={j_1}+1 \\ \{P_l,O\}}}^N \cos\theta_l\Big]\Big[\prod_{\substack{l={j_1}+1 \\ \{P_l,P_{j_1}O\}}}^N \cos\theta_l\Big] \Big[\prod_{l={j_1}+1}^N (1+\epsilon_l^{P_{j_1}O})\Big] \\
    \cdot &\, \delta_{\{P_m,O\}} \sin^2\theta_m \Big[\prod_{\substack{l=m+1 \\ \{P_l,O\}}}^{{j_1}-1} \cos^2\theta_l\Big]\Big[\prod_{l=m+1}^{{j_1}} (1+\epsilon_l^{O})\Big]\Big[\prod_{\substack{l=1 \\ \{P_l,P_mO\}}}^{m-1} \cos^2\theta_l\Big]\Big[\prod_{l=1}^{m} (1+\epsilon_l^{P_mO})\Big] \cdot [P_{j_1}O]
\end{split}
\end{equation}
In both cases $m=1,\ldots,j_1-1$, and it is easy to see that the second lines in the equations correspond to $\Omega$ factors with a $P_jO$ (respectively $O$) baseline. Finally, a slightly more involved class of diagrams is that of single bubble \textit{across} steps. When these form on $\mathcal{Z}_{j_1}$, we get a contribution of the form (see Fig.~\ref{fig:pjo-branch}c)
\begin{equation}
\begin{split}
    \substack{\Omega_m \\ \mathcal{Z}_{j_1} }[O] = i\delta_{\{P_m,O\}}\delta_{\{P_{j_1},P_mO\}}\delta_{\{P_m,P_{j_1}O\}} \cos\theta_{j_1} \sin\theta_{j_1} \Big[\prod_{\substack{l=m+1 \\ \{P_l,O\}}}^N \cos\theta_l\Big]\Big[\prod_{\substack{l=m+1 \\ \{P_l,P_{j_1}O\}}}^N \cos\theta_l\Big]\Big[\prod_{\substack{l=1 \\ \{P_l,P_{j_1}P_mO\}}}^{j_1-1} \cos^2\theta_l\Big]
    \\ \cdot (-1)^{\langle P_{j_1},P_m\rangle}\sin^2\theta_m \Big[\prod_{\substack{l=j_1+1 \\ \{P_l,P_mO\}}}^{m-1} \cos\theta_l\Big]\Big[\prod_{\substack{l=j_1+1 \\ \{P_l,P_{j_1}P_mO\}}}^{m-1} \cos\theta_l\Big]\Big[\prod_{l=1}^{m} (1+\epsilon_l^{P_{j_1}P_mO})\Big]\Big[\prod_{l=m+1}^{N} (1+\epsilon_l^{P_{j_1}O})\Big] \cdot [P_{j_1}O] 
\end{split}
\end{equation}
where $m=j_1+1,\ldots,N$. Notice that the constraint $\delta_{\{P_m,P_{j_1}O\}}$, which ensures that the bubble can close, is a reduced version of the actual constraint $\delta_{\{P_m,P_{j_1}P_mO\}}$: the two are equivalent since $P_m$ certainly commutes with itself. Furthermore, the sign factor $(-1)^{\langle P_{j_1},P_m\rangle}$ arises from the identity $P_mP_{j_1}P_mO = (-1)^{\langle P_{j_1},P_m\rangle}P_{j_1}O$ that one can use, after the bubble closes, to simplify the back-propagated observable. This is already accounted for in the last noise factor, where $\epsilon_l^{P_{j_1}O}$ is used instead of $\epsilon_l^{P_mP_{j_1}P_mO}$ (we also make use of the fact that the $\epsilon$ coefficients are insensitive to changes of sign in the operators they refer to). Interestingly, the three anti-commutation constraints immediately imply that $\langle P_{j_1},P_m\rangle = 0$, i.e., the diagram can only form if $P_{j_1}$ and $P_m$ commute. We will therefore neglect the sign factor from now on. Similarly, the $\cos\theta_{j_1}$ factor is included unconditionally, even if, strictly speaking, it would only be present if $\langle P_{j_1},P_{j_1}P_mO\rangle = 1$. In fact, such condition is equivalent to the constraint $\delta_{\{P_{j_1},P_mO\}}$. With analogous considerations, the conjugate diagram for $\mathcal{Z}'_{j_1}$ yields (see Fig.~\ref{fig:pjo-branch}d)
\begin{equation}
\begin{split}
    \substack{\Omega_m \\ \mathcal{Z}'_{j_1} }[O] = -i \delta_{\{P_m,O\}}\delta_{\{P_m,P_{j_1}O\}}\delta_{\{P_{j_1},P_mO\}}\cos\theta_{j_1} \sin\theta_{j_1}  \Big[\prod_{\substack{l=m+1 \\ \{P_l,O\}}}^N \cos\theta_l\Big] \Big[\prod_{\substack{l={j_1}+1 \\ \{P_l,P_mO\}}}^{m-1} \cos\theta_l\Big] \Big[\prod_{\substack{l=1 \\ \{P_l,P_mO\}}}^{{j_1}-1} \cos^2\theta_l\Big] \\
    \cdot \sin^2\theta_m \Big[\prod_{\substack{l={j_1}+1 \\ \{P_l,P_{j_1}P_mO\}}}^{m-1} \cos\theta_l\Big] \Big[\prod_{\substack{l=m+1 \\ \{P_l,P_{j_1}O\}}}^{N} \cos\theta_l\Big]\Big[\prod_{l=1}^{{j_1}} (1+\epsilon_l^{P_mO})\Big]\Big[\prod_{l={j_1}+1}^{m} (1+\epsilon_l^{P_{j_1}P_mO})\Big]\Big[\prod_{l=m+1}^{N} (1+\epsilon_l^{P_{j_1}O})\Big] \cdot [P_{j_1}O]
\end{split}
\end{equation}
for $m=j_1+1,\ldots,N$. Overall, we then have
\begin{equation}
    c_{j_1} = \mathcal{Z}_{j_1} + \mathcal{Z}'_{j_1} + \sum_{m=1}^{j_1-1} \Big(\Omega_m[\mathcal{Z}_{j_1}] + \mathcal{Z}'_{j_1}[\Omega_m]\Big) + \sum_{m=j_1 +1}^N \Big(\substack{\Omega_m \\ \mathcal{Z}_{j_1} } + \substack{\Omega_m \\ \mathcal{Z}'_{j_1} } \Big).
\end{equation}
To proceed with the perturbative expansion and re-summation, we will make use of the usual replacement $\cos^2\theta \mapsto 1 - \sin^2\theta$, as well as the Taylor series approximation
\begin{equation}
    \cos\theta = \sqrt{1-\sin^2\theta} \simeq 1 - \frac{1}{2}\sin^2\theta - \frac{1}{8}\sin^4\theta + \dots
    \label{eq:cos-Taylor}
\end{equation}
for all the $\cos\theta_l$ and $\cos^2\theta_l$ terms, while we will leave the global $\cos\theta_{j_1} \sin\theta_{j_1}$ factors unchanged. Since at most one virtual bubble is allowed for $k=1$ at $\kappa=4$, we will truncate the Taylor series in Eq.~\eqref{eq:cos-Taylor} at order $\sin^2\theta$. Effectively, each $\cos\theta_l$ will then spawn a virtual half-bubble (i.e., a virtual bubble with a $1/2$ coefficient). As an example, we can write
\begin{equation}
\begin{split}
    \mathcal{Z}_{j_1}
        \simeq & \, i\delta_{\{P_{j_1},O\}} \cos\theta_{j_1} \sin\theta_{j_1} 
          \Big[1+\sum_{k=1}^N \epsilon_k^{P_jO}\Big] \\
          \cdot & \, \Big[1-\frac{1}{2}\sum_{l={j_1}+1}^N\delta_{\{P_l,O\}}\sin^2\theta_l - \frac{1}{2}\sum_{l={j_1}+1}^N\delta_{\{P_l,P_{j_1}O\}}\sin^2\theta_l - \sum_{l=1}^{{j_1}-1}\delta_{\{P_l,P_{j_1}O\}}\sin^2\theta_l\Big],
\end{split}
\end{equation}
which contains two virtual half-bubble terms and one full one. With some lengthy algebraic manipulations, we can eventually derive the desired result
\begin{equation}
\begin{split}
        c_{j_1} \simeq i\cos\theta_{j_1}\sin\theta_{j_1} \bigg[ 
        \sum_{m=1}^{{j_1}-1} \sin^2\theta_m\Big( &\, 
        \dl{P_{j_1}}{O}\dl{P_m}{P_{j_1}O}\sum_{l=1}^m(\epsilon_l^{P_mP_{j_1}O}-\epsilon_l^{P_{j_1}O}) \\ &\, + \dl{P_{j_1}}{O}\dl{P_m}{O}\sum_{l=1}^m(-\epsilon_l^{P_mO}+\epsilon_l^{O})
        \Big) + \\
        + \sum_{m={j_1}+1}^{N} \sin^2\theta_m \Big( &\, \dl{P_m}{O}\dl{P_m}{P_{j_1}O}\dl{P_{j_1}}{P_mO}\sum_{l=1}^{j_1}(\epsilon_l^{P_{j_1}P_mO}-\epsilon_l^{P_mO}) \\ &\,  + \dl{P_{j_1}}{O}(\frac{1}{2}\dl{P_m}{O}+\frac{1}{2}\dl{P_m}{P_{j_1}O})\sum_{l=1}^{j_1}(-\epsilon_l^{P_{j_1}O}+\epsilon_l^{O})\Big)
        \bigg].
\end{split}
\end{equation}

\clearpage

\end{document}